\documentclass[aps,pra,reprint,twocoulumn,superscriptaddress]{revtex4-1}

\makeatletter
\AtBeginDocument{\let\LS@rot\@undefined}
\makeatother
\usepackage{pdfpages} 

\usepackage{graphicx}
\usepackage{graphics}
\usepackage{mathtools}
\usepackage{float}
\usepackage{amssymb}

\usepackage[normalem]{ulem}
\makeatletter
\def\squiggly{\bgroup \markoverwith{\textcolor{red}{\lower3.5\p@\hbox{\sixly \char58}}}\ULon}
\makeatother

\usepackage{soul,xcolor}
\usepackage[section]{placeins}
\setstcolor{red}
\definecolor{OliveGreen}{rgb}{0,0.5,0}

\usepackage[colorlinks=true,citecolor=blue,linkcolor=blue,urlcolor=blue]{hyperref}

\makeatletter
\newcommand\colorwave[1][blue]{\bgroup \markoverwith{\lower3.5\p@\hbox{\sixly \textcolor{#1}{\char58}}}\ULon}
\font\sixly=lasy6 
\makeatother

\begin{document}

\title{Static impurity in a mesoscopic system of SU(\textit{N}) fermionic matter-waves}

\begin{abstract}

We investigate the effects of a static impurity, modeled by a localized barrier, in a one-dimensional mesoscopic system comprised of strongly correlated repulsive SU($N$)-symmetric fermions. For a mesoscopic sized ring under the effect of an artificial gauge field, we analyze the energy spectrum, the particle density and the current flowing through the impurity at varying interaction strengths, barrier heights, and number of components.
%
%
%
%
%
%
We find that the physics
of the system is governed
by the competition between  effective single-particle process  and the formation of a high-stiffness spin-correlated state associated to the phenomenon of fractionalization of the flux quantum characterizing the $N$-component fermionic system.
Our findings provide a route to  probe the response of SU($N$) fermions to effective magnetic fields; at the same time,   they hold significance for fundamental understanding of localized impurity problems. 

\end{abstract}

\author{Juan Polo}
\affiliation{Quantum Research Center, Technology Innovation Institute, Abu Dhabi 9639, UAE}
\date{\today}
\author{Wayne J. Chetcuti}
\affiliation{Quantum Research Center, Technology Innovation Institute, Abu Dhabi 9639, UAE}
\author{Anna Minguzzi}
\affiliation{Université Grenoble Alpes, CNRS, LPMMC, 38000 Grenoble, France}
\author{Andreas Osterloh}
\affiliation{Quantum Research Center, Technology Innovation Institute, Abu Dhabi 9639, UAE}
\author{Luigi Amico}
\affiliation{Quantum Research Center, Technology Innovation Institute, Abu Dhabi 9639, UAE}
\affiliation{Dipartimento di Fisica e Astronomia ``Ettore Majorana'' University of Catania, Via S. Sofia 64, 95123 Catania, Italy}
\affiliation{INFN-Sezione di Catania, Via S. Sofia 64, 95123 Catania, Italy}
\maketitle


{\textit{Introduction} --} The interplay between localized impurities and correlations in quantum many-body systems is an important topic for both basic and applied physical science, ranging from mesoscopic physics~\cite{altshuler2012mesoscopic} and nano-electronics~\cite{timp1991quantum} to high-$T_c$ superconductivity~\cite{alloul2009defects} and spin liquids~\cite{kolezhuk2006theory}. 
In this regard, Kane and Fisher carried out a groundbreaking work where they investigated an interacting electronic system at low energy, and confined in an infinitely long wire interrupted by a single localized barrier~\cite{kane1992transport,kane1992transmission}. Focusing on repulsively interacting particles, even arbitrary small barriers were found to tend to infinity (under renormalization group flow).  
Such remarkable results triggered a series of studies that shed light on different aspects of strongly correlated matter
~\cite{von1998bosonization,saleur2002lectures,giamarchi2003quantum,kolezhuk2006theory,rylands2016quantum}.  

The emergence of quantum technology has ushered in a new stage of addressing impurity problems, marked by the ability to manipulate systems on a fundamental level using newly engineered physical platforms such as cold atoms and superconducting networks. This unprecedented control enables the exploration of such systems in the presence of impurities with remarkable flexibility and precision over relevant parameters, including the characteristics of barriers and the nature of particle correlations~\cite{manju2018quantum,leger2019observation,puertas2019tunable,kuzmin2021inelastic,mistakidis2022physics,leger2023revealing}. Concurrently, novel applications have arisen, leveraging the interplay between impurity and correlations to craft quantum devices with enhanced performances, spanning from Josephson junction-based devices~\cite{nadeem2023superconducting,hriscu2011coulomb,trahms2023diode,ryu2013experimental,aghamalyan2015coherent,valtolina2015josephson,singh2024shapiro} to rotation sensors~\cite{ryu2020quantum,adeniji2024double}, and interferometers wherein the static impurity can serve as a matter-wave beam splitter~\cite{godun2001prospects,haine2018quantum,wales2020splitting,naldesi2023massive}.

Our study focuses on a localized impurity in a one-dimensional mesoscopic system of $N$-component fermions, which are particles with $N$ internal degrees of freedom that can be treated as an effective spin.  On increasing the number of components, the Pauli exclusion principle relaxes allowing $N$ particles to occupy the same quantum state, with interactions effects expected to be enhanced, leading to novel and interesting physics~\cite{sonderhouse2020thermodynamics}. We note that as $N\rightarrow\infty$, keeping the number of particles $N_p<N$,  provides a ``bosonic" limit~\cite{frahm1995on}.

Specifically, we consider SU($N$) symmetric fermions with component independent repulsive interactions~\cite{gorshkov2010two,cazalilla2014ultracold,capponi2016phases}. Besides other approaches 
~\cite{kane1992transmission,kane1992transport,von1998bosonization,saleur2002lectures,giamarchi2003quantum,rylands2016quantum}, the problem can be investigated through persistent currents flowing in annular matter-wave circuits pierced by an effective magnetic field. The latter has been pursued in circuits of spinless bosons~\cite{cominotti2014optimal,aghamalyan2015coherent,polo2022quantum}. However, the spin degrees of freedom provide a significant complexity in impurity physics. When confined in mesoscopic ring-shaped potentials, at strong interactions SU($N$) fermions sustain persistent currents with fractional flux quanta $\phi_{0}/N_{p}$, with $\phi_{0}$ being the bare flux quantum of the free fermion case~\cite{chetcuti2022persistent,chetcuti2023persistent,patu2022temperature}. This phenomenon of fractionalization can be visualized 
as the formation of a collective state in which $N_p$ particles are arranged on
the ring with a high stiffness conferred by spin correlations. We refer to the latter   as `ring droplet'. This behaviour, combined with the knowledge that spin correlations and symmetries are known to play a crucial role in impurity problems~\cite{andrei1983solution}, provides a compelling motivation for our study. We note that SU($N$) fermions are experimentally realizable with alkaline earth-like cold atoms~\cite{taie2012mott,pagano2014a,scazza2014observation,hofrichterdirect2016,sonderhouse2020thermodynamics,taie2022observation}. A recent study has proposed a protocol to experimentally realize SU($N$)-symmetric systems through shielded ultracold molecules, which would circumvent the limitations of lack of tunable interactions through Feshbach resonances and different particle statistics~\cite{mukherjee2024sun,mukherjee2024sunb}.

In this paper, we show that the underlying physics of SU($N$) matter-waves flowing through a barrier is characterized by the interplay between its strength and the particle's interaction, with unique features arising from the $N$-component fermionic particles. 
Specifically: {\it i)}
we find a  distinctive response of the system to the presence of the impurity with an 
SU($N$)-dependent energy gap formation mechanism. 
{\it ii)} we demonstrate that 
impurity screening  results from the competition between 
an effective  single-particle  behavior (combined effect  of interaction and  impurity)
and the unavoidable collective phenomenon implied by the formation of the aforementioned ring droplet; {\it iii)} the impurity screening displays a  marked dependence
on the effective magnetic field through the particle density.
To support the above statements we analyze the energy spectrum of the system, particle's density and persistent current that we study through a  combination of numerical methods (exact diagonalization) and analytical techniques (Bethe ansatz) when applicable. Discussions on the physical implications of our findings are drawn in the conclusions. 
%



{\textit{SU(N) fermionic matter-wave currents} --} Consider $N_{p}$ strongly interacting $N$-component fermions of mass $m$ residing on a mesoscopic one-dimensional ring-shaped optical lattice composed of $N_{s}$ sites threaded by an effective magnetic flux $\phi$. The ring contains an impurity in the form of a localized potential barrier, breaking the discrete translational invariance. Such a scenario can be modeled  through the multi-component Fermi-Hubbard Hamiltonian~\cite{cazalilla2014ultracold,capponi2016phases}
\begin{align}\label{eq:Ham}
    \mathcal{H} = \sum\limits_{j=1}^{N_{s}}\bigg[-t\sum\limits_{\alpha}^{N}(e^{\mathrm{i}\Omega}c_{j,\alpha}^{\dagger}c_{j+1,\alpha} + e^{-\mathrm{i\Omega}}c_{j+1,\alpha}^{\dagger}c_{j,\alpha}) \nonumber \\
    + \sum\limits_{\alpha\neq\beta}^{N}U_{\alpha\beta}n_{j,\alpha}n_{j,\beta} + \sum\limits_{\alpha}^{N}\lambda_{j,\alpha}n_{j,\alpha}\bigg ],
\end{align}
where $c_{j,\alpha}^{\dagger}$ ($c_{j,\alpha}$) creates (destroys) a fermion with colour $\alpha$ on site $j$ and $n_{j,\alpha}=c_{j,\alpha}^{\dagger}c_{j,\alpha}$ is the local number operator. The parameters $t$ and $U$ correspond to the hopping and on-site interaction energies respectively. Unless explicitly stated, we adopt isotropic interactions $U_{\alpha\beta} = U$, turning Eq.~\eqref{eq:Ham} into the SU($N$) Hubbard model. In this case: (i) it is a model that well represents experimentally realizable lattices of alkaline earth-like cold atoms~\cite{scazza2014observation,cazalilla2014ultracold}; (ii) the system is, for $\lambda_{\alpha}=0$, Bethe Ansatz integrable in the continuous limit of vanishing lattice spacing as it tends to the Gaudin-Yang-Sutherland model~\cite{sutherland1968further,chetcuti2023probe}. The barrier of strength $\lambda$ is equal for all colours localized at site $j_{0}$ such that $\lambda_{j,\alpha} = \lambda\delta_{j,j_{0}}$, taken to be positive. All energies will be given in units of $t$ and we consider an equal number of particles per component.

The artificial gauge field $\Omega = (2\pi\phi)/(N_{s}\phi_{0})$ is introduced through the Peierls substitution $t\rightarrow t e^{\mathrm{i}\Omega}$~\cite{peierls1933zur}. For neutral cold atoms, the synthetic field can be introduced through various means~\cite{amico2021roadmap}, resulting in a Hamiltonian of the same form as that in Eq.~\eqref{eq:Ham} albeit with a different parametric expression for the elementary flux quantum $\phi_{0}$, which encodes the physical nature of the specific implementation. One instance is that of inducing rotation through stirring~\cite{wright2013driving,polo2024persistent}, which is particularly suitable in the presence of a barrier, where $\phi_{0} = \hbar/(mR^{2})$ with $R$ being the radius of the ring (taken to be equal to 1 in our case). In the presence of a flux, the many-body spectrum in the free particle regime is piece-wise parabolic, caused by energy level crossings between parabolas of well-defined angular momentum per particle, denoted by $\ell$, to counteract the increase in flux piercing the system. Consequently, the energy spectrum is periodic in $\phi$ with a period fixed by $\phi_{0}$ akin to that of particles in a periodic potential. Therefore, following Leggett, $E(\phi)$ define the `Bloch bands' of the problem in which the magnetic flux plays the role of the momentum~\cite{leggett1991dephasing}.  

In order to characterize the flow of SU($N$) fermionic particles through a localized barrier, we utilize the spatial density $\langle n_j \rangle$ and the persistent current $I(\phi)$, which is the system's response to the applied field $\phi$. At zero temperature, the persistent current is obtained from the  ground-state energy $E_{0}$: by the relation
$I(\phi)=-\partial E_{0}/\partial\phi$. We employ the Hellmann-Feynman theorem to calculate the species-wise current $I_{\alpha}(\phi)$, which for lattice systems, reads
$I_{\alpha}(\phi) = -(2\mathrm{i} t\pi)/(\phi_{0}N_s)\sum_{j=1}^{N_{s}}\langle e^{\mathrm{i}\Omega}c_{j,\alpha}^{\dagger}c_{j+1,\alpha}-\mathrm{h.c.}\rangle_{GS}$
with $\langle\bullet\rangle_{GS}$ being the ground-state expectation value. The resulting persistent current profile is a saw-tooth shape with the jumps corresponding to changes in the system's angular momentum due to level crossings. In cold atoms implementations, the saw-tooth character of the current gives rise to the discrete steps in the angular momentum per particle as observed in experiments~\cite{polo2024persistent}. 

Fermionic matter-wave currents, without an impurity, in the strongly interacting regime exhibit a reduced periodicity depending on the nature of interactions the particles are subjected to. For repulsive interactions, the phenomenon arises from energy crossings between the ground-state and higher energy excitations characterized by different spin quantum numbers. 
Such a feature leads to an energy landscape consisting of $N_{p}$ piece-wise parabolic segments per flux quantum~\cite{chetcuti2022persistent}.   (see~\cite{chetcuti2023probe} for attractive case). The reduced periodicity of the current reflects a fractional flux quantum  $\phi_{0}/N_{p}$ mediated by energy level crossings at $n\phi_{0}/(2N_{p})$ for odd integer $n$. The phenomenon of fractionalization is specific to a mesoscopic system with periodic boundary conditions reflecting the formation of a collective state, dubbed ring droplet,  in which  the specific spin correlations give rise to  a stiff particle arrangement ~\cite{chetcuti2022persistent}.  



{\textit{SU($N$) dependent spectral gaps formation --}} The presence of an impurity in a single-component quantum system, including bosons with repulsive or attractive interactions, is known to split \textit{all} the degeneracies in $E_{0}(\phi)$ opening a spectral gap $\Delta$ ~\cite{aghamalyan2015coherent,cominotti2014optimal,polo2022quantum,polo2024persistent}.  At these avoided crossings, a coherent superposition of different angular momentum states is created. Consequently, the persistent current's saw-tooth shape starts to smoothen out at weak barriers, eventually becoming a sinusoid as $\lambda$ gets large enough. 
\begin{figure}[h!]
    \centering
    \includegraphics[width=1\linewidth]{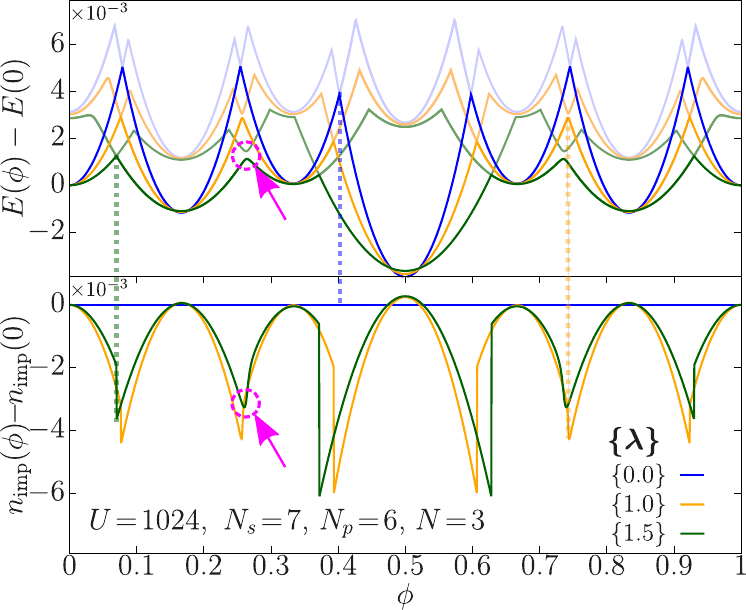}
    \put(-15,193){(\textbf{a})}
    \put(-15,102){(\textbf{b})}
    \caption{
    Profiles of the energy $E(\phi)$ (top) and corresponding density at the impurity site $n_{\mathrm{imp}}$ (bottom) against the effective magnetic flux $\phi$ in the regime of strong interactions $U/t = 1024$ for various barrier strengths $\lambda/t$ in a ring of $N_{s}$ sites. 
    Dotted lines highlight the connection between the energy peaks and density minima, whilst circles indicate the opening of the gaps in $E(\phi)$ and designated changes in $n_{\mathrm{imp}}$.
    %
    The quadratic Casimir values $s$ for each energy parabola are $\{6,3,3,0,3,3,6\}$. Results obtained with exact diagonalization of the SU($N$) Hubbard model using the parameter set indicated in the figure. 
    }
    \label{fig:gap}
\end{figure}

For the present case of interacting SU($N$) fermions with an impurity, only specific degeneracies in $E_{0}(\phi)$ are found to split. The mechanism behind these selective gap openings lies in the internal degrees of freedom~\footnote{On account of this, such behaviour is different than that of bosons, even the ones with attractive interactions where fractionalization is present.}. Indeed, as the barrier in Eq.~\eqref{eq:Ham} commutes with the quadratic SU($N$) Casimir operators with eigenvalue $s$, which characterizes each piece-wise parabola of the energy landscape (see Supplemental material;~\cite{osterloh2023exact,pecci2023persistent}), it is unable to couple states with different values of $s$\footnote{Note that the SU($N$) representation can be different, which would generally be indicated by a different $s$. However, for some $N$ the different representations can share the same $s$.}. In turn, the Bloch bands and the  corresponding persistent current landscape are found to display a non-trivial dependence on $s$. Specifically, depending on $N_{p}$ and $N$, the ground and excited states intersect each other as $U$ increases: $N_{p}$ dictates the amount of energy level crossings; 
$N$ governs the quantity $\&$  $\phi$-dependence of the energy gaps through the SU($N$) Casimir. 
Such behavior reflects the dependence of the system’s ground-state on $s$. 
Consequently, only crossings characterized by the same Casimir value can be split by the barrier to generate a gap -- Fig.~\ref{fig:gap}. 

For weak barriers, we find that $\Delta$ scales as $\lambda/U$ (see Supplemental material). 
The $1/U$ behaviour can be understood by a  perturbative calculation around the Bethe Ansatz results $\lambda=0$  as $U\rightarrow\infty$~\cite{ogata1990bethe,osterloh2023exact}, which  gives the energy contribution of the spin part scaling  as  $t^{2}/U$ \cite{osterloh2023exact}.  
For the case of stronger barriers, we find that the gap scales non-linearly with $\lambda$, as $\displaystyle{\Delta (U,\lambda)\approx \frac{\lambda^{\gamma}}{U}}$ with $\gamma>1$ outside the linear response behaviour expected at low $\lambda$.

We note that impurity and fractionalization compete with each other.
{In particular, the opening of the gap  makes the energy landscape 
'protected' from fractionalization 
as soon interactions are strong enough to overcome 
the gap $\Delta$ between the ground state and the  higher excitations energy  (characterized by different spin quantum numbers). 
Therefore,  the low lying spectrum at weak interaction in the presence of the impurity is not affected by the correlations emerging due to flux quantum fractionalization. In this sense, such regime  is effectively single particle.}   
%



{\textit{Screening of the local impurity} --} 
Due to the barrier's presence, there is a global minimum in the particle's density at the site where it is located, denoted by $n_{\mathrm{imp}}$, meaning that particles tend to avoid residing in the barrier to minimize $E(\phi)$. 
The reduced value of $n_{\mathrm{imp}}$ with decreasing $N_{p}/N$ reflects the loosening of the Pauli principle 
(enabling more particles to reside at the impurity site which enhances the barrier - see Supplemental material). 
The increase of $n_{\mathrm{imp}}$  with the effective interaction $N U$, signals that the fluid screens the barrier. 
For weak $U$, the rate at which the barrier is screened $\partial n_{imp}/\partial U$ is larger with increasing $N$
~\footnote{The barrier is the least effective versus interactions at $N_{p}=N$. In this case, the density at the impurity site coincides with that of a bosonic system of the same $N_{p}$. Such a behaviour reflects the lack of a meaningful Pauli exclusion principle in the system.}. In contrast,  $\partial n_{imp}/\partial U$ decreases with $N$ for intermediate and large $U$, with $n_{\mathrm{imp}}$ saturating  to a value, coinciding with that of hard-core bosons/spinless fermions, that is independent of  $N$ as $U\rightarrow\infty$. We note that when  $N_{s}$ is comparable to $N_{p}$, the density at the barrier is found to be non-monotonic with  $U$ -- see Supplemental material.
Indeed, such specific screening properties arise because the presence of the impurity sets the aforementioned   effectively weak interaction regime 
in which the spin  correlations implied by the  flux-quantum fractionalization  are not effective. At larger interactions, instead, the flux-quantum fractionalization is relevant, and the properties of the system are dominated  by the formation of the `ring droplet' of $N_{p}$ particles. 

Besides capturing the screening effect, the density profile can also provide information about the current when a barrier is present. In particular, Fig.~\ref{fig:gap}(b) shows the flux dependence of the density reflecting both the energy landscape and the current in the system.
As particles increase their flow, the barrier counteracts this motion by reducing the density at the impurity increasing its effectiveness. When the flow starts to reduce, i.e. when the groundstate energy decreases, the density at the impurity increases again reaching a maximum at zero current. We also note that smoothened gaps and energy crossings correspond to smooth and discontinuous jumps of $n_{imp}(\phi)$, which is also reflected in the density-density correlations (see Supplemental).

\begin{figure}[h!]
    \centering
    \includegraphics[width=1\linewidth]{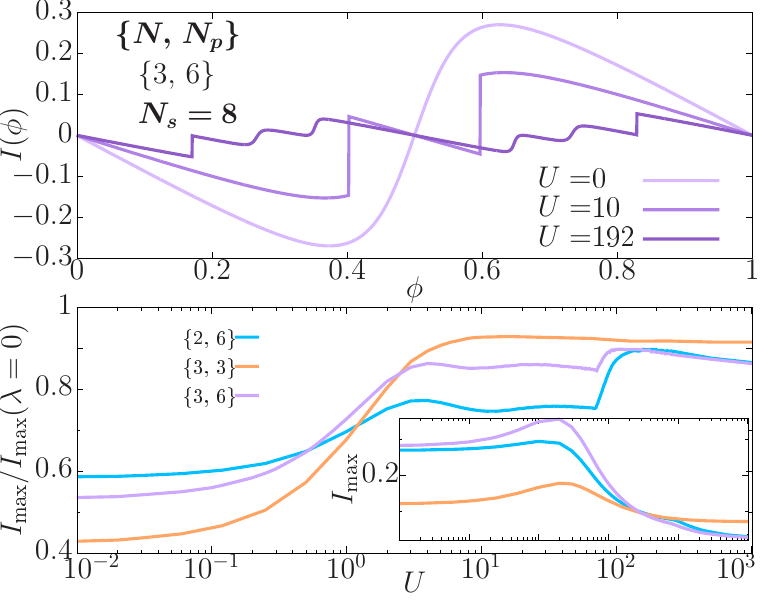}
    \put(-220,185){(\textbf{a})}
    \put(-220,90){(\textbf{b})}
    \caption{
    (a) Current $I(\phi)$ versus flux $\phi$ showcasing the interplay between the characteristic fractionalized sawtooth and a smoothened profile given by the presence of the impurity.
    (b) Maximum persistent current amplitude $I_{\mathrm{max}}$ as a function of interactions $U/t$ in the presence of a barrier with strength $\lambda/t=1$ for different number of particles $N_{p}$ and components $N$. In the main panel he current is normalized by the maximum current computed at $\lambda=0$ at all values of interactions, while inset shows the bare value of $I_\text{max}$ versus $U$.
    The competition between screening of the barrier and fractionalization results in $I_\text{max}U^{\nu}$, with $\nu\approx 0.7$ for $U/t\ll1$ and $\nu\approx -0.8$ for $U/t\gg1$).
    Results obtained with exact diagonalization of the SU($N$) Hubbard model.}
    \label{fig:Imax_vs_U}
\end{figure}

{\textit{Persistent current through a localized barrier --}} The interplay between the impurity strength and interaction manifests itself in the persistent current profile. Focusing on the current's maximum amplitude $I_{\mathrm{max}}$, we find that it exhibits a non-monotonic behaviour as a function of $\lambda$ and $U$ -- Fig.~\ref{fig:Imax_vs_U}. 
Such behavior arises because of the aforementioned competition between the 
effective single-particle current, leading to the screening of the impurity and thus to a current increase, and the formation of the ring droplet,  resulting in the suppression of the current 
(the ring droplet  leads to an effective increase of the mass of the particles.)
%
In the following, we analyze the above competition as function of interaction $U$ and number of fermion components $N$.
For  weak interactions, and fixed $\lambda$, the current amplitude increases with $U$  at a rate that is more pronounced with increasing $N$ --  a similar  behavior occurs in the density $n_{imp}$  (see Supplemental material). 
We note that $N_p/N=1$ leads to a vanishing  Pauli principle constraint.  In this case, and at weak interactions, the bosonic limit of persistent currents is recovered~\cite{cominotti2014optimal}. For intermediate interactions such that $U/\lambda > 1$, where the phenomenon of fractionalization becomes more pronounced, the $I_{\mathrm{max}}$ reflects the impurity screening with a clear dependence on $U$. 
In this regime the current undergoes an `hybridization' becoming a smoothed yet fractionalized one with cusps, exhibiting a reduced periodicity (see Supplemental material). Subsequently, $I_{\mathrm{max}}$ emerges from the cusped parabolas. 
We point out that in this regime, even for $N_p/N=1$, our persistent current is markedly different from that of single-component bosons  - this feature is natural since in the latter system, no fractionalization occurs~\cite{polo2024persistent}. Such difference is clearly showcased in Fig.\ref{fig:Imax_vs_U}(b) in which the persistent current displays additional maxima and specific cusps that are a direct consequence of the fractionalization phenomenon (occurring because of specific level crossings implied by fractionalizaiton).

In the limit of strong interactions, as $U/\lambda\gg 1$, the impact of the impurity is drastically reduced. Nonetheless, we find that the current is strongly suppressed. Such an effect primarily originates from the fractionalization whose rate is enhanced with $N$~\cite{chetcuti2022persistent}. Consistently, the persistent current displays a perfect saw-tooth shape with a reduced periodicity of $1/N_{p}$ exhibiting negligible signs of smearing. 







{\textit{Discussions and Conclusions} --} We considered a  single localized impurity embedded in a one-dimensional mesoscopic system of repulsive SU($N$) fermions subjected to an artificial gauge field with flux $\phi$. 
 The physics is characterized by  the interplay between the impurity strength and interaction through the phenomenon of  flux quantum fractionalization that is a genuine feature of multi-component systems~\cite{chetcuti2022persistent}.
Density-density correlations corroborate the two different regimes: for weak interactions the correlations change gradually, signaling the effective single-particle behavior, while in the strongly interacting regime, a bunching effect occurs, meaning that correlations in the ring droplet change discontinuously only for persistent currents with different winding numbers. 

Below, we summarize our results for the energy spectrum of the system,  density at the impurity site $n_{imp}$ and persistent current $I(\phi)$.
 
 The impurity opens energy gaps $\Delta\sim 1/U$ {\it selectively},  following an SU($N$) dependent mechanism - see Fig.\ref{fig:gap}(a). 
At weak interactions,
the flux-quantum-fractionalization has a small influence on the impurity physics 
as long as the gap to the higher energy excitations is smaller than  $\Delta$. 
In such a regime, the system is characterized by an  effective single-particle spectrum landscape (renormalized by the interaction). For increasing interactions, the fractionalization of the flux-quantum becomes relevant 
and signals the
formation of a ring droplet.



By looking at $n_{imp}$ as a function of $U$, we can establish that initially the barrier is screened with a rate dependence that becomes larger with $N$:   As more particles can reside at the impurity site, the impurity is screened faster since the  effective repulsion is enhanced as $NU$ in SU($N$) systems. 
As $U\rightarrow\infty$, $n_{imp}$ is found to saturate to a value that is independent of $N$ because of the  formation  of the  ring droplet with a stiff particle arrangement~\cite{chetcuti2023persistent}. 
Besides capturing the screening effect, {\it $n_{imp}$ displays a marked dependence on  $\phi$}, reflecting the periodicity of the SU($N$) fermions flux quantum fractionalization - see Fig.\ref{fig:gap}(b). In a cold atom systems, $n_{imp}$ would be accessible, for example, through in-situ density measurements (see \cite{litvinov2021measuring} for a recent work on a $1d$ fermionic gas).  
Such a feature 
can be exploited to access to the phenomenon of fractionalization in cold atoms experiments beyond the  interference dynamics~\cite{chetcuti2023interference}.
We note that the latter, being related to the momenta of particles and therefore to the current, is also found affected by the impurity. 


%

As a result of the aforementioned competition between effective single-particle behavior and  ring droplet formation, the current amplitude {\it $I_{\mathrm{max}}$ exhibits a non-monotonic behaviour} as a function of the interaction and parametrized by the impurity strength $\lambda$. 
Whilst at weak $U$, $I_{\mathrm{max}}$ increases due to the suppression of the impurity as observed in the density $n_{imp}$, this is not the case on going to stronger interactions. In this regime, the  current  decreases as the result of the  spin correlations  characterizing the ring droplet. Such distinctive features of SU($N$) matter-wave emerge in Fig.\ref{fig:Imax_vs_U}. 

The study of SU($N$) fermions confined in rings poses an interesting challenge both from an analytical and numerical standpoint. Analytical techniques have been found at strong interactions \cite{osterloh2023exact} while 
numerically
one faces known problems coming from 
periodic boundary conditions (as e.g. in DMRG) combined with a high degeneracy in the ground-state. 
To address these limits, further development are being explored \cite{botzung2024exact,weichselbaum2024qspace}. On the analytical side, we remark that our study is in a very different regime in which the standard one-dimensional Luttinger theory, as the Fermi sphere in our system is hardly filled~\cite{giamarchi2003quantum} nonetheless, we expect to obtain similar features. As shown in our results and as predicted by Bethe Ansatz\cite{chetcuti2022persistent},  the fractionalization is a direct consequence of the coupling between the spin and 'charges' rapidities. However, Luttinger theory implements spin-charge separation, with the caveat that the two sectors in the Luttinger Hamiltonian (in the mesoscopic regime) are still coupled through certain constraints on spin and charge quantum numbers \cite{seidelLuther2005}; as a result, 
the charge and spin current operators $J_c$ and $J_s$ respectively are functionally related.
Because of the latter feature, the factorization shows up in terms of energy crossings  in the energy response to  Aharonov–Bohm flux $\phi$ $E_c(J_c;\phi)\propto (J_c - \phi/\phi_0)^2$ (see also~\cite{kusmartsev1994aharonov}). Moreover, the selective gap openings we discussed in our manuscript holds true also in the Luttinger liquid picture. 
In this context, a full renormalization group description~\cite{gogolin2004bosonization} including the barrier effects would  provide an interesting step forward.

We highlight that static impurities are  of high  relevance for quantum technology. We note that  the  experimental realization based on alkali-earth atoms  that was  proposed recently for rings with a  spin impurity\cite{amaricci2025engineering} can be pursued for the present case with nearly the same ingredients.
 The setup considered here can also provide the basis  for current-based simulators and and interferometers~\cite{amico2021roadmap,amico2022colloquium,polo2024perspective,naldesi2022enhancing,polo2022quantum,naldesi2023massive} utilizing $N$-component matter-waves.


{\textit{Acknowledgements} --} We thank Enrico C. Domanti and Sam Carr for useful discussions.

\bibliography{refs}

\begin{thebibliography}{71}%
\makeatletter
\providecommand \@ifxundefined [1]{%
 \@ifx{#1\undefined}
}%
\providecommand \@ifnum [1]{%
 \ifnum #1\expandafter \@firstoftwo
 \else \expandafter \@secondoftwo
 \fi
}%
\providecommand \@ifx [1]{%
 \ifx #1\expandafter \@firstoftwo
 \else \expandafter \@secondoftwo
 \fi
}%
\providecommand \natexlab [1]{#1}%
\providecommand \enquote  [1]{``#1''}%
\providecommand \bibnamefont  [1]{#1}%
\providecommand \bibfnamefont [1]{#1}%
\providecommand \citenamefont [1]{#1}%
\providecommand \href@noop [0]{\@secondoftwo}%
\providecommand \href [0]{\begingroup \@sanitize@url \@href}%
\providecommand \@href[1]{\@@startlink{#1}\@@href}%
\providecommand \@@href[1]{\endgroup#1\@@endlink}%
\providecommand \@sanitize@url [0]{\catcode `\\12\catcode `\$12\catcode
  `\&12\catcode `\#12\catcode `\^12\catcode `\_12\catcode `\%12\relax}%
\providecommand \@@startlink[1]{}%
\providecommand \@@endlink[0]{}%
\providecommand \url  [0]{\begingroup\@sanitize@url \@url }%
\providecommand \@url [1]{\endgroup\@href {#1}{\urlprefix }}%
\providecommand \urlprefix  [0]{URL }%
\providecommand \Eprint [0]{\href }%
\providecommand \doibase [0]{http://dx.doi.org/}%
\providecommand \selectlanguage [0]{\@gobble}%
\providecommand \bibinfo  [0]{\@secondoftwo}%
\providecommand \bibfield  [0]{\@secondoftwo}%
\providecommand \translation [1]{[#1]}%
\providecommand \BibitemOpen [0]{}%
\providecommand \bibitemStop [0]{}%
\providecommand \bibitemNoStop [0]{.\EOS\space}%
\providecommand \EOS [0]{\spacefactor3000\relax}%
\providecommand \BibitemShut  [1]{\csname bibitem#1\endcsname}%
\let\auto@bib@innerbib\@empty
\bibitem [{\citenamefont {Altshuler}\ \emph {et~al.}(2012)\citenamefont
  {Altshuler}, \citenamefont {Lee},\ and\ \citenamefont
  {Webb}}]{altshuler2012mesoscopic}%
  \BibitemOpen
  \bibfield  {author} {\bibinfo {author} {\bibfnamefont {B.~L.}\ \bibnamefont
  {Altshuler}}, \bibinfo {author} {\bibfnamefont {P.~A.}\ \bibnamefont {Lee}},
  \ and\ \bibinfo {author} {\bibfnamefont {W.~R.}\ \bibnamefont {Webb}},\
  }\href@noop {} {\emph {\bibinfo {title} {Mesoscopic phenomena in solids}}}\
  (\bibinfo  {publisher} {Elsevier},\ \bibinfo {year} {2012})\BibitemShut
  {NoStop}%
\bibitem [{\citenamefont {Timp}\ and\ \citenamefont
  {Howard}(1991)}]{timp1991quantum}%
  \BibitemOpen
  \bibfield  {author} {\bibinfo {author} {\bibfnamefont {G.~L.}\ \bibnamefont
  {Timp}}\ and\ \bibinfo {author} {\bibfnamefont {R.~E.}\ \bibnamefont
  {Howard}},\ }\href@noop {} {\bibfield  {journal} {\bibinfo  {journal} {IEEE
  Proc.}\ }\textbf {\bibinfo {volume} {79}},\ \bibinfo {pages} {1188} (\bibinfo
  {year} {1991})}\BibitemShut {NoStop}%
\bibitem [{\citenamefont {Alloul}\ \emph {et~al.}(2009)\citenamefont {Alloul},
  \citenamefont {Bobroff}, \citenamefont {Gabay},\ and\ \citenamefont
  {Hirschfeld}}]{alloul2009defects}%
  \BibitemOpen
  \bibfield  {author} {\bibinfo {author} {\bibfnamefont {H.}~\bibnamefont
  {Alloul}}, \bibinfo {author} {\bibfnamefont {J.}~\bibnamefont {Bobroff}},
  \bibinfo {author} {\bibfnamefont {M.}~\bibnamefont {Gabay}}, \ and\ \bibinfo
  {author} {\bibfnamefont {P.}~\bibnamefont {Hirschfeld}},\ }\href@noop {}
  {\bibfield  {journal} {\bibinfo  {journal} {Rev. Mod. Phys.}\ }\textbf
  {\bibinfo {volume} {81}},\ \bibinfo {pages} {45} (\bibinfo {year}
  {2009})}\BibitemShut {NoStop}%
\bibitem [{\citenamefont {Kolezhuk}\ \emph {et~al.}(2006)\citenamefont
  {Kolezhuk}, \citenamefont {Sachdev}, \citenamefont {Biswas},\ and\
  \citenamefont {Chen}}]{kolezhuk2006theory}%
  \BibitemOpen
  \bibfield  {author} {\bibinfo {author} {\bibfnamefont {A.}~\bibnamefont
  {Kolezhuk}}, \bibinfo {author} {\bibfnamefont {S.}~\bibnamefont {Sachdev}},
  \bibinfo {author} {\bibfnamefont {R.~R.}\ \bibnamefont {Biswas}}, \ and\
  \bibinfo {author} {\bibfnamefont {P.}~\bibnamefont {Chen}},\ }\href@noop {}
  {\bibfield  {journal} {\bibinfo  {journal} {Phys. Rev. B}\ }\textbf {\bibinfo
  {volume} {74}},\ \bibinfo {pages} {165114} (\bibinfo {year}
  {2006})}\BibitemShut {NoStop}%
\bibitem [{\citenamefont {Kane}\ and\ \citenamefont
  {Fisher}(1992{\natexlab{a}})}]{kane1992transport}%
  \BibitemOpen
  \bibfield  {author} {\bibinfo {author} {\bibfnamefont {C.}~\bibnamefont
  {Kane}}\ and\ \bibinfo {author} {\bibfnamefont {M.~P.}\ \bibnamefont
  {Fisher}},\ }\href@noop {} {\bibfield  {journal} {\bibinfo  {journal} {Phys.
  Rev. Lett.}\ }\textbf {\bibinfo {volume} {68}},\ \bibinfo {pages} {1220}
  (\bibinfo {year} {1992}{\natexlab{a}})}\BibitemShut {NoStop}%
\bibitem [{\citenamefont {Kane}\ and\ \citenamefont
  {Fisher}(1992{\natexlab{b}})}]{kane1992transmission}%
  \BibitemOpen
  \bibfield  {author} {\bibinfo {author} {\bibfnamefont {C.}~\bibnamefont
  {Kane}}\ and\ \bibinfo {author} {\bibfnamefont {M.~P.}\ \bibnamefont
  {Fisher}},\ }\href@noop {} {\bibfield  {journal} {\bibinfo  {journal}
  {Physical Review B}\ }\textbf {\bibinfo {volume} {46}},\ \bibinfo {pages}
  {15233} (\bibinfo {year} {1992}{\natexlab{b}})}\BibitemShut {NoStop}%
\bibitem [{\citenamefont {Von~Delft}\ and\ \citenamefont
  {Schoeller}(1998)}]{von1998bosonization}%
  \BibitemOpen
  \bibfield  {author} {\bibinfo {author} {\bibfnamefont {J.}~\bibnamefont
  {Von~Delft}}\ and\ \bibinfo {author} {\bibfnamefont {H.}~\bibnamefont
  {Schoeller}},\ }\href@noop {} {\bibfield  {journal} {\bibinfo  {journal}
  {Annalen der Physik}\ }\textbf {\bibinfo {volume} {510}},\ \bibinfo {pages}
  {225} (\bibinfo {year} {1998})}\BibitemShut {NoStop}%
\bibitem [{\citenamefont {Saleur}(2002)}]{saleur2002lectures}%
  \BibitemOpen
  \bibfield  {author} {\bibinfo {author} {\bibfnamefont {H.}~\bibnamefont
  {Saleur}},\ }in\ \href@noop {} {\emph {\bibinfo {booktitle} {Topological
  aspects of low dimensional systems: Session LXIX}}}\ (\bibinfo  {publisher}
  {Springer},\ \bibinfo {year} {2002})\ p.\ \bibinfo {pages} {473}\BibitemShut
  {NoStop}%
\bibitem [{\citenamefont {Giamarchi}(2003)}]{giamarchi2003quantum}%
  \BibitemOpen
  \bibfield  {author} {\bibinfo {author} {\bibfnamefont {T.}~\bibnamefont
  {Giamarchi}},\ }\href@noop {} {\emph {\bibinfo {title} {Quantum physics in
  one dimension}}},\ Vol.\ \bibinfo {volume} {121}\ (\bibinfo  {publisher}
  {Clarendon Press},\ \bibinfo {year} {2003})\BibitemShut {NoStop}%
\bibitem [{\citenamefont {Rylands}\ and\ \citenamefont
  {Andrei}(2016)}]{rylands2016quantum}%
  \BibitemOpen
  \bibfield  {author} {\bibinfo {author} {\bibfnamefont {C.}~\bibnamefont
  {Rylands}}\ and\ \bibinfo {author} {\bibfnamefont {N.}~\bibnamefont
  {Andrei}},\ }\href@noop {} {\bibfield  {journal} {\bibinfo  {journal} {Phys.
  Rev. B}\ }\textbf {\bibinfo {volume} {94}},\ \bibinfo {pages} {115142}
  (\bibinfo {year} {2016})}\BibitemShut {NoStop}%
\bibitem [{\citenamefont {Manju}\ \emph {et~al.}(2018)\citenamefont {Manju},
  \citenamefont {Hardman}, \citenamefont {Sooriyabandara}, \citenamefont
  {Wigley}, \citenamefont {Close}, \citenamefont {Robins}, \citenamefont
  {Hush},\ and\ \citenamefont {Szigeti}}]{manju2018quantum}%
  \BibitemOpen
  \bibfield  {author} {\bibinfo {author} {\bibfnamefont {P.}~\bibnamefont
  {Manju}}, \bibinfo {author} {\bibfnamefont {K.}~\bibnamefont {Hardman}},
  \bibinfo {author} {\bibfnamefont {M.}~\bibnamefont {Sooriyabandara}},
  \bibinfo {author} {\bibfnamefont {P.}~\bibnamefont {Wigley}}, \bibinfo
  {author} {\bibfnamefont {J.}~\bibnamefont {Close}}, \bibinfo {author}
  {\bibfnamefont {N.}~\bibnamefont {Robins}}, \bibinfo {author} {\bibfnamefont
  {M.}~\bibnamefont {Hush}}, \ and\ \bibinfo {author} {\bibfnamefont
  {S.}~\bibnamefont {Szigeti}},\ }\href@noop {} {\bibfield  {journal} {\bibinfo
   {journal} {Physical Review A}\ }\textbf {\bibinfo {volume} {98}},\ \bibinfo
  {pages} {053629} (\bibinfo {year} {2018})}\BibitemShut {NoStop}%
\bibitem [{\citenamefont {L{\'e}ger}\ \emph {et~al.}(2019)\citenamefont
  {L{\'e}ger}, \citenamefont {Puertas-Mart{\'\i}nez}, \citenamefont
  {Bharadwaj}, \citenamefont {Dassonneville}, \citenamefont {Delaforce},
  \citenamefont {Foroughi}, \citenamefont {Milchakov}, \citenamefont {Planat},
  \citenamefont {Buisson}, \citenamefont {Naud} \emph
  {et~al.}}]{leger2019observation}%
  \BibitemOpen
  \bibfield  {author} {\bibinfo {author} {\bibfnamefont {S.}~\bibnamefont
  {L{\'e}ger}}, \bibinfo {author} {\bibfnamefont {J.}~\bibnamefont
  {Puertas-Mart{\'\i}nez}}, \bibinfo {author} {\bibfnamefont {K.}~\bibnamefont
  {Bharadwaj}}, \bibinfo {author} {\bibfnamefont {R.}~\bibnamefont
  {Dassonneville}}, \bibinfo {author} {\bibfnamefont {J.}~\bibnamefont
  {Delaforce}}, \bibinfo {author} {\bibfnamefont {F.}~\bibnamefont {Foroughi}},
  \bibinfo {author} {\bibfnamefont {V.}~\bibnamefont {Milchakov}}, \bibinfo
  {author} {\bibfnamefont {L.}~\bibnamefont {Planat}}, \bibinfo {author}
  {\bibfnamefont {O.}~\bibnamefont {Buisson}}, \bibinfo {author} {\bibfnamefont
  {C.}~\bibnamefont {Naud}},  \emph {et~al.},\ }\href@noop {} {\bibfield
  {journal} {\bibinfo  {journal} {Nat. Comm.}\ }\textbf {\bibinfo {volume}
  {10}},\ \bibinfo {pages} {5259} (\bibinfo {year} {2019})}\BibitemShut
  {NoStop}%
\bibitem [{\citenamefont {Puertas~Mart{\'\i}nez}\ \emph
  {et~al.}(2019)\citenamefont {Puertas~Mart{\'\i}nez}, \citenamefont
  {L{\'e}ger}, \citenamefont {Gheeraert}, \citenamefont {Dassonneville},
  \citenamefont {Planat}, \citenamefont {Foroughi}, \citenamefont {Krupko},
  \citenamefont {Buisson}, \citenamefont {Naud}, \citenamefont
  {Hasch-Guichard}, \citenamefont {Florense}, \citenamefont {Snyman},\ and\
  \citenamefont {Roch}}]{puertas2019tunable}%
  \BibitemOpen
  \bibfield  {author} {\bibinfo {author} {\bibfnamefont {J.}~\bibnamefont
  {Puertas~Mart{\'\i}nez}}, \bibinfo {author} {\bibfnamefont {S.}~\bibnamefont
  {L{\'e}ger}}, \bibinfo {author} {\bibfnamefont {N.}~\bibnamefont
  {Gheeraert}}, \bibinfo {author} {\bibfnamefont {R.}~\bibnamefont
  {Dassonneville}}, \bibinfo {author} {\bibfnamefont {L.}~\bibnamefont
  {Planat}}, \bibinfo {author} {\bibfnamefont {F.}~\bibnamefont {Foroughi}},
  \bibinfo {author} {\bibfnamefont {Y.}~\bibnamefont {Krupko}}, \bibinfo
  {author} {\bibfnamefont {O.}~\bibnamefont {Buisson}}, \bibinfo {author}
  {\bibfnamefont {C.}~\bibnamefont {Naud}}, \bibinfo {author} {\bibfnamefont
  {W.}~\bibnamefont {Hasch-Guichard}}, \bibinfo {author} {\bibfnamefont
  {S.}~\bibnamefont {Florense}}, \bibinfo {author} {\bibfnamefont
  {I.}~\bibnamefont {Snyman}}, \ and\ \bibinfo {author} {\bibfnamefont
  {N.}~\bibnamefont {Roch}},\ }\href@noop {} {\bibfield  {journal} {\bibinfo
  {journal} {New Phys. J. Q. Inf.}\ }\textbf {\bibinfo {volume} {5}},\ \bibinfo
  {pages} {19} (\bibinfo {year} {2019})}\BibitemShut {NoStop}%
\bibitem [{\citenamefont {Kuzmin}\ \emph {et~al.}(2021)\citenamefont {Kuzmin},
  \citenamefont {Grabon}, \citenamefont {Mehta}, \citenamefont {Burshtein},
  \citenamefont {Goldstein}, \citenamefont {Houzet}, \citenamefont {Glazman},\
  and\ \citenamefont {Manucharyan}}]{kuzmin2021inelastic}%
  \BibitemOpen
  \bibfield  {author} {\bibinfo {author} {\bibfnamefont {R.}~\bibnamefont
  {Kuzmin}}, \bibinfo {author} {\bibfnamefont {N.}~\bibnamefont {Grabon}},
  \bibinfo {author} {\bibfnamefont {N.}~\bibnamefont {Mehta}}, \bibinfo
  {author} {\bibfnamefont {A.}~\bibnamefont {Burshtein}}, \bibinfo {author}
  {\bibfnamefont {M.}~\bibnamefont {Goldstein}}, \bibinfo {author}
  {\bibfnamefont {M.}~\bibnamefont {Houzet}}, \bibinfo {author} {\bibfnamefont
  {L.~I.}\ \bibnamefont {Glazman}}, \ and\ \bibinfo {author} {\bibfnamefont
  {V.~E.}\ \bibnamefont {Manucharyan}},\ }\href@noop {} {\bibfield  {journal}
  {\bibinfo  {journal} {Phys. Rev. Lett.}\ }\textbf {\bibinfo {volume} {126}},\
  \bibinfo {pages} {197701} (\bibinfo {year} {2021})}\BibitemShut {NoStop}%
\bibitem [{\citenamefont {Mistakidis}\ and\ \citenamefont
  {Volosniev}(2022)}]{mistakidis2022physics}%
  \BibitemOpen
  \bibfield  {author} {\bibinfo {author} {\bibfnamefont {S.}~\bibnamefont
  {Mistakidis}}\ and\ \bibinfo {author} {\bibfnamefont {A.}~\bibnamefont
  {Volosniev}},\ }\href@noop {} {\emph {\bibinfo {title} {Physics of Impurities
  in Quantum Gases}}}\ (\bibinfo  {publisher} {MDPI},\ \bibinfo {year}
  {2022})\BibitemShut {NoStop}%
\bibitem [{\citenamefont {L{\'e}ger}\ \emph {et~al.}(2023)\citenamefont
  {L{\'e}ger}, \citenamefont {S{\'e}pulcre}, \citenamefont {Fraudet},
  \citenamefont {Buisson}, \citenamefont {Naud}, \citenamefont
  {Hasch-Guichard}, \citenamefont {Florens}, \citenamefont {Snyman},
  \citenamefont {Basko},\ and\ \citenamefont {Roch}}]{leger2023revealing}%
  \BibitemOpen
  \bibfield  {author} {\bibinfo {author} {\bibfnamefont {S.}~\bibnamefont
  {L{\'e}ger}}, \bibinfo {author} {\bibfnamefont {T.}~\bibnamefont
  {S{\'e}pulcre}}, \bibinfo {author} {\bibfnamefont {D.}~\bibnamefont
  {Fraudet}}, \bibinfo {author} {\bibfnamefont {O.}~\bibnamefont {Buisson}},
  \bibinfo {author} {\bibfnamefont {C.}~\bibnamefont {Naud}}, \bibinfo {author}
  {\bibfnamefont {W.}~\bibnamefont {Hasch-Guichard}}, \bibinfo {author}
  {\bibfnamefont {S.}~\bibnamefont {Florens}}, \bibinfo {author} {\bibfnamefont
  {I.}~\bibnamefont {Snyman}}, \bibinfo {author} {\bibfnamefont {D.~M.}\
  \bibnamefont {Basko}}, \ and\ \bibinfo {author} {\bibfnamefont
  {N.}~\bibnamefont {Roch}},\ }\href@noop {} {\bibfield  {journal} {\bibinfo
  {journal} {SciPost Phys.}\ }\textbf {\bibinfo {volume} {14}},\ \bibinfo
  {pages} {130} (\bibinfo {year} {2023})}\BibitemShut {NoStop}%
\bibitem [{\citenamefont {Nadeem}\ \emph {et~al.}(2023)\citenamefont {Nadeem},
  \citenamefont {Fuhrer},\ and\ \citenamefont
  {Wang}}]{nadeem2023superconducting}%
  \BibitemOpen
  \bibfield  {author} {\bibinfo {author} {\bibfnamefont {M.}~\bibnamefont
  {Nadeem}}, \bibinfo {author} {\bibfnamefont {M.~S.}\ \bibnamefont {Fuhrer}},
  \ and\ \bibinfo {author} {\bibfnamefont {X.}~\bibnamefont {Wang}},\
  }\href@noop {} {\bibfield  {journal} {\bibinfo  {journal} {Nature Rev.
  Phys.}\ }\textbf {\bibinfo {volume} {5}},\ \bibinfo {pages} {558} (\bibinfo
  {year} {2023})}\BibitemShut {NoStop}%
\bibitem [{\citenamefont {Hriscu}\ and\ \citenamefont
  {Nazarov}(2011)}]{hriscu2011coulomb}%
  \BibitemOpen
  \bibfield  {author} {\bibinfo {author} {\bibfnamefont {A.}~\bibnamefont
  {Hriscu}}\ and\ \bibinfo {author} {\bibfnamefont {Y.~V.}\ \bibnamefont
  {Nazarov}},\ }\href@noop {} {\bibfield  {journal} {\bibinfo  {journal} {Phys.
  Rev. B}\ }\textbf {\bibinfo {volume} {83}},\ \bibinfo {pages} {174511}
  (\bibinfo {year} {2011})}\BibitemShut {NoStop}%
\bibitem [{\citenamefont {Trahms}\ \emph {et~al.}(2023)\citenamefont {Trahms},
  \citenamefont {Melischek}, \citenamefont {Steiner}, \citenamefont {Mahendru},
  \citenamefont {Tamir}, \citenamefont {Bogdanoff}, \citenamefont {Peters},
  \citenamefont {Reecht}, \citenamefont {Winkelmann}, \citenamefont {von Oppen}
  \emph {et~al.}}]{trahms2023diode}%
  \BibitemOpen
  \bibfield  {author} {\bibinfo {author} {\bibfnamefont {M.}~\bibnamefont
  {Trahms}}, \bibinfo {author} {\bibfnamefont {L.}~\bibnamefont {Melischek}},
  \bibinfo {author} {\bibfnamefont {J.~F.}\ \bibnamefont {Steiner}}, \bibinfo
  {author} {\bibfnamefont {B.}~\bibnamefont {Mahendru}}, \bibinfo {author}
  {\bibfnamefont {I.}~\bibnamefont {Tamir}}, \bibinfo {author} {\bibfnamefont
  {N.}~\bibnamefont {Bogdanoff}}, \bibinfo {author} {\bibfnamefont
  {O.}~\bibnamefont {Peters}}, \bibinfo {author} {\bibfnamefont
  {G.}~\bibnamefont {Reecht}}, \bibinfo {author} {\bibfnamefont {C.~B.}\
  \bibnamefont {Winkelmann}}, \bibinfo {author} {\bibfnamefont
  {F.}~\bibnamefont {von Oppen}},  \emph {et~al.},\ }\href@noop {} {\bibfield
  {journal} {\bibinfo  {journal} {Nature}\ }\textbf {\bibinfo {volume} {615}},\
  \bibinfo {pages} {628} (\bibinfo {year} {2023})}\BibitemShut {NoStop}%
\bibitem [{\citenamefont {Ryu}\ \emph {et~al.}(2013)\citenamefont {Ryu},
  \citenamefont {Blackburn}, \citenamefont {Blinova},\ and\ \citenamefont
  {Boshier}}]{ryu2013experimental}%
  \BibitemOpen
  \bibfield  {author} {\bibinfo {author} {\bibfnamefont {C.}~\bibnamefont
  {Ryu}}, \bibinfo {author} {\bibfnamefont {P.~W.}\ \bibnamefont {Blackburn}},
  \bibinfo {author} {\bibfnamefont {A.~A.}\ \bibnamefont {Blinova}}, \ and\
  \bibinfo {author} {\bibfnamefont {M.~G.}\ \bibnamefont {Boshier}},\ }\href
  {\doibase 10.1103/PhysRevLett.111.205301} {\bibfield  {journal} {\bibinfo
  {journal} {Phys. Rev. Lett.}\ }\textbf {\bibinfo {volume} {111}},\ \bibinfo
  {pages} {205301} (\bibinfo {year} {2013})}\BibitemShut {NoStop}%
\bibitem [{\citenamefont {Aghamalyan}\ \emph {et~al.}(2015)\citenamefont
  {Aghamalyan}, \citenamefont {Cominotti}, \citenamefont {Rizzi}, \citenamefont
  {Rossini}, \citenamefont {Hekking}, \citenamefont {Minguzzi}, \citenamefont
  {Kwek},\ and\ \citenamefont {Amico}}]{aghamalyan2015coherent}%
  \BibitemOpen
  \bibfield  {author} {\bibinfo {author} {\bibfnamefont {D.}~\bibnamefont
  {Aghamalyan}}, \bibinfo {author} {\bibfnamefont {M.}~\bibnamefont
  {Cominotti}}, \bibinfo {author} {\bibfnamefont {M.}~\bibnamefont {Rizzi}},
  \bibinfo {author} {\bibfnamefont {D.}~\bibnamefont {Rossini}}, \bibinfo
  {author} {\bibfnamefont {F.}~\bibnamefont {Hekking}}, \bibinfo {author}
  {\bibfnamefont {A.}~\bibnamefont {Minguzzi}}, \bibinfo {author}
  {\bibfnamefont {L.-C.}\ \bibnamefont {Kwek}}, \ and\ \bibinfo {author}
  {\bibfnamefont {L.}~\bibnamefont {Amico}},\ }\href {\doibase
  10.1088/1367-2630/17/4/045023} {\bibfield  {journal} {\bibinfo  {journal}
  {New J. Phys.}\ }\textbf {\bibinfo {volume} {17}},\ \bibinfo {pages} {045023}
  (\bibinfo {year} {2015})}\BibitemShut {NoStop}%
\bibitem [{\citenamefont {Valtolina}\ \emph {et~al.}(2015)\citenamefont
  {Valtolina}, \citenamefont {Burchianti}, \citenamefont {Amico}, \citenamefont
  {Neri}, \citenamefont {Xhani}, \citenamefont {Seman}, \citenamefont
  {Trombettoni}, \citenamefont {Smerzi}, \citenamefont {Zaccanti},
  \citenamefont {Inguscio},\ and\ \citenamefont
  {Roati}}]{valtolina2015josephson}%
  \BibitemOpen
  \bibfield  {author} {\bibinfo {author} {\bibfnamefont {G.}~\bibnamefont
  {Valtolina}}, \bibinfo {author} {\bibfnamefont {A.}~\bibnamefont
  {Burchianti}}, \bibinfo {author} {\bibfnamefont {A.}~\bibnamefont {Amico}},
  \bibinfo {author} {\bibfnamefont {E.}~\bibnamefont {Neri}}, \bibinfo {author}
  {\bibfnamefont {K.}~\bibnamefont {Xhani}}, \bibinfo {author} {\bibfnamefont
  {J.~A.}\ \bibnamefont {Seman}}, \bibinfo {author} {\bibfnamefont
  {A.}~\bibnamefont {Trombettoni}}, \bibinfo {author} {\bibfnamefont
  {A.}~\bibnamefont {Smerzi}}, \bibinfo {author} {\bibfnamefont
  {M.}~\bibnamefont {Zaccanti}}, \bibinfo {author} {\bibfnamefont
  {M.}~\bibnamefont {Inguscio}}, \ and\ \bibinfo {author} {\bibfnamefont
  {G.}~\bibnamefont {Roati}},\ }\href@noop {} {\bibfield  {journal} {\bibinfo
  {journal} {Science}\ }\textbf {\bibinfo {volume} {350}},\ \bibinfo {pages}
  {1505} (\bibinfo {year} {2015})}\BibitemShut {NoStop}%
\bibitem [{\citenamefont {Singh}\ \emph {et~al.}(2024)\citenamefont {Singh},
  \citenamefont {Polo}, \citenamefont {Mathey},\ and\ \citenamefont
  {Amico}}]{singh2024shapiro}%
  \BibitemOpen
  \bibfield  {author} {\bibinfo {author} {\bibfnamefont {V.~P.}\ \bibnamefont
  {Singh}}, \bibinfo {author} {\bibfnamefont {J.}~\bibnamefont {Polo}},
  \bibinfo {author} {\bibfnamefont {L.}~\bibnamefont {Mathey}}, \ and\ \bibinfo
  {author} {\bibfnamefont {L.}~\bibnamefont {Amico}},\ }\href@noop {}
  {\bibfield  {journal} {\bibinfo  {journal} {Physical Review Letters}\
  }\textbf {\bibinfo {volume} {133}},\ \bibinfo {pages} {093401} (\bibinfo
  {year} {2024})}\BibitemShut {NoStop}%
\bibitem [{\citenamefont {Ryu}\ \emph {et~al.}(2020)\citenamefont {Ryu},
  \citenamefont {Samson},\ and\ \citenamefont {Boshier}}]{ryu2020quantum}%
  \BibitemOpen
  \bibfield  {author} {\bibinfo {author} {\bibfnamefont {C.}~\bibnamefont
  {Ryu}}, \bibinfo {author} {\bibfnamefont {E.~C.}\ \bibnamefont {Samson}}, \
  and\ \bibinfo {author} {\bibfnamefont {M.~G.}\ \bibnamefont {Boshier}},\
  }\href {\doibase 10.1038/s41467-020-17185-6} {\bibfield  {journal} {\bibinfo
  {journal} {Nat. Comm.}\ }\textbf {\bibinfo {volume} {11}} (\bibinfo {year}
  {2020}),\ 10.1038/s41467-020-17185-6}\BibitemShut {NoStop}%
\bibitem [{\citenamefont {Adeniji}\ \emph {et~al.}(2024)\citenamefont
  {Adeniji}, \citenamefont {Henry}, \citenamefont {Thomas}, \citenamefont
  {Sapp}, \citenamefont {Goyal}, \citenamefont {Clark},\ and\ \citenamefont
  {Edwards}}]{adeniji2024double}%
  \BibitemOpen
  \bibfield  {author} {\bibinfo {author} {\bibfnamefont {O.}~\bibnamefont
  {Adeniji}}, \bibinfo {author} {\bibfnamefont {C.}~\bibnamefont {Henry}},
  \bibinfo {author} {\bibfnamefont {S.}~\bibnamefont {Thomas}}, \bibinfo
  {author} {\bibfnamefont {R.~C.}\ \bibnamefont {Sapp}}, \bibinfo {author}
  {\bibfnamefont {A.}~\bibnamefont {Goyal}}, \bibinfo {author} {\bibfnamefont
  {C.~W.}\ \bibnamefont {Clark}}, \ and\ \bibinfo {author} {\bibfnamefont
  {M.}~\bibnamefont {Edwards}},\ }\href {https://arxiv.org/abs/2411.06585}
  {\enquote {\bibinfo {title} {{Double-target BEC atomtronic rotation
  sensor}},}\ } (\bibinfo {year} {2024}),\ \Eprint
  {http://arxiv.org/abs/2411.06585} {arXiv:2411.06585 [cond-mat.quant-gas]}
  \BibitemShut {NoStop}%
\bibitem [{\citenamefont {Godun}\ \emph {et~al.}(2001)\citenamefont {Godun},
  \citenamefont {d'Arcy}, \citenamefont {Summy},\ and\ \citenamefont
  {Burnett}}]{godun2001prospects}%
  \BibitemOpen
  \bibfield  {author} {\bibinfo {author} {\bibfnamefont {R.}~\bibnamefont
  {Godun}}, \bibinfo {author} {\bibfnamefont {M.}~\bibnamefont {d'Arcy}},
  \bibinfo {author} {\bibfnamefont {G.}~\bibnamefont {Summy}}, \ and\ \bibinfo
  {author} {\bibfnamefont {K.}~\bibnamefont {Burnett}},\ }\href@noop {}
  {\bibfield  {journal} {\bibinfo  {journal} {Contemp. Phys.}\ }\textbf
  {\bibinfo {volume} {42}},\ \bibinfo {pages} {77} (\bibinfo {year}
  {2001})}\BibitemShut {NoStop}%
\bibitem [{\citenamefont {Haine}(2018)}]{haine2018quantum}%
  \BibitemOpen
  \bibfield  {author} {\bibinfo {author} {\bibfnamefont {S.~A.}\ \bibnamefont
  {Haine}},\ }\href@noop {} {\bibfield  {journal} {\bibinfo  {journal} {New J.
  Phys.}\ }\textbf {\bibinfo {volume} {20}},\ \bibinfo {pages} {033009}
  (\bibinfo {year} {2018})}\BibitemShut {NoStop}%
\bibitem [{\citenamefont {Wales}\ \emph {et~al.}(2020)\citenamefont {Wales},
  \citenamefont {Rakonjac}, \citenamefont {Billam}, \citenamefont {Helm},
  \citenamefont {Gardiner},\ and\ \citenamefont
  {Cornish}}]{wales2020splitting}%
  \BibitemOpen
  \bibfield  {author} {\bibinfo {author} {\bibfnamefont {O.~J.}\ \bibnamefont
  {Wales}}, \bibinfo {author} {\bibfnamefont {A.}~\bibnamefont {Rakonjac}},
  \bibinfo {author} {\bibfnamefont {T.~P.}\ \bibnamefont {Billam}}, \bibinfo
  {author} {\bibfnamefont {J.~L.}\ \bibnamefont {Helm}}, \bibinfo {author}
  {\bibfnamefont {S.~A.}\ \bibnamefont {Gardiner}}, \ and\ \bibinfo {author}
  {\bibfnamefont {S.~L.}\ \bibnamefont {Cornish}},\ }\href@noop {} {\bibfield
  {journal} {\bibinfo  {journal} {Communications Physics}\ }\textbf {\bibinfo
  {volume} {3}},\ \bibinfo {pages} {51} (\bibinfo {year} {2020})}\BibitemShut
  {NoStop}%
\bibitem [{\citenamefont {Naldesi}\ \emph {et~al.}(2023)\citenamefont
  {Naldesi}, \citenamefont {Polo}, \citenamefont {Drummond}, \citenamefont
  {Dunjko}, \citenamefont {Amico}, \citenamefont {Minguzzi},\ and\
  \citenamefont {Olshanii}}]{naldesi2023massive}%
  \BibitemOpen
  \bibfield  {author} {\bibinfo {author} {\bibfnamefont {P.}~\bibnamefont
  {Naldesi}}, \bibinfo {author} {\bibfnamefont {J.}~\bibnamefont {Polo}},
  \bibinfo {author} {\bibfnamefont {P.~D.}\ \bibnamefont {Drummond}}, \bibinfo
  {author} {\bibfnamefont {V.}~\bibnamefont {Dunjko}}, \bibinfo {author}
  {\bibfnamefont {L.}~\bibnamefont {Amico}}, \bibinfo {author} {\bibfnamefont
  {A.}~\bibnamefont {Minguzzi}}, \ and\ \bibinfo {author} {\bibfnamefont
  {M.}~\bibnamefont {Olshanii}},\ }\href {\doibase
  10.21468/SciPostPhys.15.5.187} {\bibfield  {journal} {\bibinfo  {journal}
  {SciPost Physics}\ }\textbf {\bibinfo {volume} {15}},\ \bibinfo {pages} {187}
  (\bibinfo {year} {2023})}\BibitemShut {NoStop}%
\bibitem [{\citenamefont {Sonderhouse}\ \emph {et~al.}(2020)\citenamefont
  {Sonderhouse}, \citenamefont {Sanner}, \citenamefont {Hutson}, \citenamefont
  {Goban}, \citenamefont {Bilitewski}, \citenamefont {Yan}, \citenamefont
  {Milner}, \citenamefont {Rey},\ and\ \citenamefont
  {Ye}}]{sonderhouse2020thermodynamics}%
  \BibitemOpen
  \bibfield  {author} {\bibinfo {author} {\bibfnamefont {L.}~\bibnamefont
  {Sonderhouse}}, \bibinfo {author} {\bibfnamefont {C.}~\bibnamefont {Sanner}},
  \bibinfo {author} {\bibfnamefont {R.~B.}\ \bibnamefont {Hutson}}, \bibinfo
  {author} {\bibfnamefont {A.}~\bibnamefont {Goban}}, \bibinfo {author}
  {\bibfnamefont {T.}~\bibnamefont {Bilitewski}}, \bibinfo {author}
  {\bibfnamefont {L.}~\bibnamefont {Yan}}, \bibinfo {author} {\bibfnamefont
  {W.~R.}\ \bibnamefont {Milner}}, \bibinfo {author} {\bibfnamefont {A.~M.}\
  \bibnamefont {Rey}}, \ and\ \bibinfo {author} {\bibfnamefont
  {J.}~\bibnamefont {Ye}},\ }\href {\doibase 10.1038/s41567-020-0986-6}
  {\bibfield  {journal} {\bibinfo  {journal} {Nature Physics}\ }\textbf
  {\bibinfo {volume} {16}},\ \bibinfo {pages} {1216} (\bibinfo {year}
  {2020})}\BibitemShut {NoStop}%
\bibitem [{\citenamefont {Frahm}\ and\ \citenamefont
  {Schadschneider}(1995)}]{frahm1995on}%
  \BibitemOpen
  \bibfield  {author} {\bibinfo {author} {\bibfnamefont {H.}~\bibnamefont
  {Frahm}}\ and\ \bibinfo {author} {\bibfnamefont {A.}~\bibnamefont
  {Schadschneider}},\ }\enquote {\bibinfo {title} {On the bethe ansatz soluble
  degenerate hubbard model},}\ in\ \href {\doibase 10.1007/978-1-4899-1042-4_2}
  {\emph {\bibinfo {booktitle} {The Hubbard Model: Its Physics and Mathematical
  Physics}}},\ \bibinfo {editor} {edited by\ \bibinfo {editor} {\bibfnamefont
  {D.}~\bibnamefont {Baeriswyl}}, \bibinfo {editor} {\bibfnamefont {D.~K.}\
  \bibnamefont {Campbell}}, \bibinfo {editor} {\bibfnamefont {J.~M.~P.}\
  \bibnamefont {Carmelo}}, \bibinfo {editor} {\bibfnamefont {F.}~\bibnamefont
  {Guinea}}, \ and\ \bibinfo {editor} {\bibfnamefont {E.}~\bibnamefont
  {Louis}}}\ (\bibinfo  {publisher} {Springer US},\ \bibinfo {address} {Boston,
  MA},\ \bibinfo {year} {1995})\ p.~\bibinfo {pages} {21}\BibitemShut {NoStop}%
\bibitem [{\citenamefont {Gorshkov}\ \emph {et~al.}(2010)\citenamefont
  {Gorshkov}, \citenamefont {Hermele}, \citenamefont {Gurarie}, \citenamefont
  {Xu}, \citenamefont {Julienne}, \citenamefont {Ye}, \citenamefont {Zoller},
  \citenamefont {Demler}, \citenamefont {Lukin},\ and\ \citenamefont
  {Rey}}]{gorshkov2010two}%
  \BibitemOpen
  \bibfield  {author} {\bibinfo {author} {\bibfnamefont {A.~V.}\ \bibnamefont
  {Gorshkov}}, \bibinfo {author} {\bibfnamefont {M.}~\bibnamefont {Hermele}},
  \bibinfo {author} {\bibfnamefont {V.}~\bibnamefont {Gurarie}}, \bibinfo
  {author} {\bibfnamefont {C.}~\bibnamefont {Xu}}, \bibinfo {author}
  {\bibfnamefont {P.~S.}\ \bibnamefont {Julienne}}, \bibinfo {author}
  {\bibfnamefont {J.}~\bibnamefont {Ye}}, \bibinfo {author} {\bibfnamefont
  {P.}~\bibnamefont {Zoller}}, \bibinfo {author} {\bibfnamefont
  {E.}~\bibnamefont {Demler}}, \bibinfo {author} {\bibfnamefont {M.~D.}\
  \bibnamefont {Lukin}}, \ and\ \bibinfo {author} {\bibfnamefont {A.~M.}\
  \bibnamefont {Rey}},\ }\href {\doibase 10.1038/nphys1535} {\bibfield
  {journal} {\bibinfo  {journal} {Nature Physics}\ }\textbf {\bibinfo {volume}
  {6}},\ \bibinfo {pages} {289} (\bibinfo {year} {2010})}\BibitemShut {NoStop}%
\bibitem [{\citenamefont {Cazalilla}\ and\ \citenamefont
  {Rey}(2014)}]{cazalilla2014ultracold}%
  \BibitemOpen
  \bibfield  {author} {\bibinfo {author} {\bibfnamefont {M.~A.}\ \bibnamefont
  {Cazalilla}}\ and\ \bibinfo {author} {\bibfnamefont {A.~M.}\ \bibnamefont
  {Rey}},\ }\href {\doibase 10.1088/0034-4885/77/12/124401} {\bibfield
  {journal} {\bibinfo  {journal} {Reports on Progress in Physics}\ }\textbf
  {\bibinfo {volume} {77}},\ \bibinfo {pages} {124401} (\bibinfo {year}
  {2014})}\BibitemShut {NoStop}%
\bibitem [{\citenamefont {Capponi}\ \emph {et~al.}(2016)\citenamefont
  {Capponi}, \citenamefont {Lecheminant},\ and\ \citenamefont
  {Totsuka}}]{capponi2016phases}%
  \BibitemOpen
  \bibfield  {author} {\bibinfo {author} {\bibfnamefont {S.}~\bibnamefont
  {Capponi}}, \bibinfo {author} {\bibfnamefont {P.}~\bibnamefont
  {Lecheminant}}, \ and\ \bibinfo {author} {\bibfnamefont {K.}~\bibnamefont
  {Totsuka}},\ }\href {\doibase https://doi.org/10.1016/j.aop.2016.01.011}
  {\bibfield  {journal} {\bibinfo  {journal} {Annals of Physics}\ }\textbf
  {\bibinfo {volume} {367}},\ \bibinfo {pages} {50} (\bibinfo {year}
  {2016})}\BibitemShut {NoStop}%
\bibitem [{\citenamefont {Cominotti}\ \emph {et~al.}(2014)\citenamefont
  {Cominotti}, \citenamefont {Rossini}, \citenamefont {Rizzi}, \citenamefont
  {Hekking},\ and\ \citenamefont {Minguzzi}}]{cominotti2014optimal}%
  \BibitemOpen
  \bibfield  {author} {\bibinfo {author} {\bibfnamefont {M.}~\bibnamefont
  {Cominotti}}, \bibinfo {author} {\bibfnamefont {D.}~\bibnamefont {Rossini}},
  \bibinfo {author} {\bibfnamefont {M.}~\bibnamefont {Rizzi}}, \bibinfo
  {author} {\bibfnamefont {F.}~\bibnamefont {Hekking}}, \ and\ \bibinfo
  {author} {\bibfnamefont {A.}~\bibnamefont {Minguzzi}},\ }\href@noop {}
  {\bibfield  {journal} {\bibinfo  {journal} {Phys. Rev. Lett.}\ }\textbf
  {\bibinfo {volume} {113}},\ \bibinfo {pages} {025301} (\bibinfo {year}
  {2014})}\BibitemShut {NoStop}%
\bibitem [{\citenamefont {Polo}\ \emph {et~al.}(2021)\citenamefont {Polo},
  \citenamefont {Naldesi}, \citenamefont {Minguzzi},\ and\ \citenamefont
  {Amico}}]{polo2022quantum}%
  \BibitemOpen
  \bibfield  {author} {\bibinfo {author} {\bibfnamefont {J.}~\bibnamefont
  {Polo}}, \bibinfo {author} {\bibfnamefont {P.}~\bibnamefont {Naldesi}},
  \bibinfo {author} {\bibfnamefont {A.}~\bibnamefont {Minguzzi}}, \ and\
  \bibinfo {author} {\bibfnamefont {L.}~\bibnamefont {Amico}},\ }\href
  {\doibase 10.1088/2058-9565/ac39f6} {\bibfield  {journal} {\bibinfo
  {journal} {Q. Sci. Tech.}\ }\textbf {\bibinfo {volume} {7}},\ \bibinfo
  {pages} {015015} (\bibinfo {year} {2021})}\BibitemShut {NoStop}%
\bibitem [{\citenamefont {Chetcuti}\ \emph {et~al.}(2022)\citenamefont
  {Chetcuti}, \citenamefont {Haug}, \citenamefont {Kwek},\ and\ \citenamefont
  {Amico}}]{chetcuti2022persistent}%
  \BibitemOpen
  \bibfield  {author} {\bibinfo {author} {\bibfnamefont {W.~J.}\ \bibnamefont
  {Chetcuti}}, \bibinfo {author} {\bibfnamefont {T.}~\bibnamefont {Haug}},
  \bibinfo {author} {\bibfnamefont {L.-C.}\ \bibnamefont {Kwek}}, \ and\
  \bibinfo {author} {\bibfnamefont {L.}~\bibnamefont {Amico}},\ }\href
  {\doibase 10.21468/SciPostPhys.12.1.033} {\bibfield  {journal} {\bibinfo
  {journal} {SciPost Physics}\ }\textbf {\bibinfo {volume} {12}},\ \bibinfo
  {pages} {33} (\bibinfo {year} {2022})}\BibitemShut {NoStop}%
\bibitem [{\citenamefont {Chetcuti}(2023)}]{chetcuti2023persistent}%
  \BibitemOpen
  \bibfield  {author} {\bibinfo {author} {\bibfnamefont {W.~J.}\ \bibnamefont
  {Chetcuti}},\ }\emph {\bibinfo {title} {{Persistent Currents in Atomtronic
  Circuits of SU($N$) Fermions}}},\ \href@noop {} {Ph.D. thesis},\ \bibinfo
  {school} {University of Catania} (\bibinfo {year} {2023}),\ \Eprint
  {http://arxiv.org/abs/2311.03072} {2311.03072} \BibitemShut {NoStop}%
\bibitem [{\citenamefont {P\^a\ifmmode~\mbox{\c{t}}\else \c{t}\fi{}u}\ and\
  \citenamefont {Averin}(2022)}]{patu2022temperature}%
  \BibitemOpen
  \bibfield  {author} {\bibinfo {author} {\bibfnamefont {O.~I.}\ \bibnamefont
  {P\^a\ifmmode~\mbox{\c{t}}\else \c{t}\fi{}u}}\ and\ \bibinfo {author}
  {\bibfnamefont {D.~V.}\ \bibnamefont {Averin}},\ }\href {\doibase
  10.1103/PhysRevLett.128.096801} {\bibfield  {journal} {\bibinfo  {journal}
  {Phys. Rev. Lett.}\ }\textbf {\bibinfo {volume} {128}},\ \bibinfo {pages}
  {096801} (\bibinfo {year} {2022})}\BibitemShut {NoStop}%
\bibitem [{\citenamefont {Andrei}\ \emph {et~al.}(1983)\citenamefont {Andrei},
  \citenamefont {Furuya},\ and\ \citenamefont
  {Lowenstein}}]{andrei1983solution}%
  \BibitemOpen
  \bibfield  {author} {\bibinfo {author} {\bibfnamefont {N.}~\bibnamefont
  {Andrei}}, \bibinfo {author} {\bibfnamefont {K.}~\bibnamefont {Furuya}}, \
  and\ \bibinfo {author} {\bibfnamefont {J.~H.}\ \bibnamefont {Lowenstein}},\
  }\href {\doibase 10.1103/RevModPhys.55.331} {\bibfield  {journal} {\bibinfo
  {journal} {Reviews of Modern Physics}\ }\textbf {\bibinfo {volume} {55}},\
  \bibinfo {pages} {331} (\bibinfo {year} {1983})}\BibitemShut {NoStop}%
\bibitem [{\citenamefont {Taie}\ \emph {et~al.}(2012)\citenamefont {Taie},
  \citenamefont {Yamazaki}, \citenamefont {Sugawa},\ and\ \citenamefont
  {Takahashi}}]{taie2012mott}%
  \BibitemOpen
  \bibfield  {author} {\bibinfo {author} {\bibfnamefont {S.}~\bibnamefont
  {Taie}}, \bibinfo {author} {\bibfnamefont {R.}~\bibnamefont {Yamazaki}},
  \bibinfo {author} {\bibfnamefont {S.}~\bibnamefont {Sugawa}}, \ and\ \bibinfo
  {author} {\bibfnamefont {Y.}~\bibnamefont {Takahashi}},\ }\href {\doibase
  10.1038/nphys2430} {\bibfield  {journal} {\bibinfo  {journal} {Nature
  Physics}\ }\textbf {\bibinfo {volume} {8}},\ \bibinfo {pages} {825} (\bibinfo
  {year} {2012})}\BibitemShut {NoStop}%
\bibitem [{\citenamefont {Pagano}\ \emph {et~al.}(2014)\citenamefont {Pagano},
  \citenamefont {Mancini}, \citenamefont {Cappellini}, \citenamefont
  {Lombardi}, \citenamefont {Sch\"{a}fer}, \citenamefont {Hu}, \citenamefont
  {Liu}, \citenamefont {Catani}, \citenamefont {Sias}, \citenamefont
  {Inguscio},\ and\ \citenamefont {Fallani}}]{pagano2014a}%
  \BibitemOpen
  \bibfield  {author} {\bibinfo {author} {\bibfnamefont {G.}~\bibnamefont
  {Pagano}}, \bibinfo {author} {\bibfnamefont {M.}~\bibnamefont {Mancini}},
  \bibinfo {author} {\bibfnamefont {G.}~\bibnamefont {Cappellini}}, \bibinfo
  {author} {\bibfnamefont {P.}~\bibnamefont {Lombardi}}, \bibinfo {author}
  {\bibfnamefont {F.}~\bibnamefont {Sch\"{a}fer}}, \bibinfo {author}
  {\bibfnamefont {H.}~\bibnamefont {Hu}}, \bibinfo {author} {\bibfnamefont
  {X.-J.}\ \bibnamefont {Liu}}, \bibinfo {author} {\bibfnamefont
  {J.}~\bibnamefont {Catani}}, \bibinfo {author} {\bibfnamefont
  {C.}~\bibnamefont {Sias}}, \bibinfo {author} {\bibfnamefont {M.}~\bibnamefont
  {Inguscio}}, \ and\ \bibinfo {author} {\bibfnamefont {L.}~\bibnamefont
  {Fallani}},\ }\href {\doibase 10.1038/nphys2878} {\bibfield  {journal}
  {\bibinfo  {journal} {Nature Physics}\ }\textbf {\bibinfo {volume} {10}},\
  \bibinfo {pages} {198} (\bibinfo {year} {2014})}\BibitemShut {NoStop}%
\bibitem [{\citenamefont {Scazza}\ \emph {et~al.}(2014)\citenamefont {Scazza},
  \citenamefont {Hofrichter}, \citenamefont {H{\"o}fer}, \citenamefont
  {De~Groot}, \citenamefont {Bloch},\ and\ \citenamefont
  {F{\"o}lling}}]{scazza2014observation}%
  \BibitemOpen
  \bibfield  {author} {\bibinfo {author} {\bibfnamefont {F.}~\bibnamefont
  {Scazza}}, \bibinfo {author} {\bibfnamefont {C.}~\bibnamefont {Hofrichter}},
  \bibinfo {author} {\bibfnamefont {M.}~\bibnamefont {H{\"o}fer}}, \bibinfo
  {author} {\bibfnamefont {P.}~\bibnamefont {De~Groot}}, \bibinfo {author}
  {\bibfnamefont {I.}~\bibnamefont {Bloch}}, \ and\ \bibinfo {author}
  {\bibfnamefont {S.}~\bibnamefont {F{\"o}lling}},\ }\href@noop {} {\bibfield
  {journal} {\bibinfo  {journal} {Nature Physics}\ }\textbf {\bibinfo {volume}
  {10}},\ \bibinfo {pages} {779} (\bibinfo {year} {2014})}\BibitemShut
  {NoStop}%
\bibitem [{\citenamefont {Hofrichter}\ \emph {et~al.}(2016)\citenamefont
  {Hofrichter}, \citenamefont {Riegger}, \citenamefont {Scazza}, \citenamefont
  {H\"ofer}, \citenamefont {Fernandes}, \citenamefont {Bloch},\ and\
  \citenamefont {F\"olling}}]{hofrichterdirect2016}%
  \BibitemOpen
  \bibfield  {author} {\bibinfo {author} {\bibfnamefont {C.}~\bibnamefont
  {Hofrichter}}, \bibinfo {author} {\bibfnamefont {L.}~\bibnamefont {Riegger}},
  \bibinfo {author} {\bibfnamefont {F.}~\bibnamefont {Scazza}}, \bibinfo
  {author} {\bibfnamefont {M.}~\bibnamefont {H\"ofer}}, \bibinfo {author}
  {\bibfnamefont {D.~R.}\ \bibnamefont {Fernandes}}, \bibinfo {author}
  {\bibfnamefont {I.}~\bibnamefont {Bloch}}, \ and\ \bibinfo {author}
  {\bibfnamefont {S.}~\bibnamefont {F\"olling}},\ }\href {\doibase
  10.1103/PhysRevX.6.021030} {\bibfield  {journal} {\bibinfo  {journal} {Phys.
  Rev. X}\ }\textbf {\bibinfo {volume} {6}},\ \bibinfo {pages} {021030}
  (\bibinfo {year} {2016})}\BibitemShut {NoStop}%
\bibitem [{\citenamefont {Taie}\ \emph {et~al.}(2022)\citenamefont {Taie},
  \citenamefont {Ibarra-Garc{\'{\i}}a-Padilla}, \citenamefont {Nishizawa},
  \citenamefont {Takasu}, \citenamefont {Kuno}, \citenamefont {Wei},
  \citenamefont {Scalettar}, \citenamefont {Hazzard},\ and\ \citenamefont
  {Takahashi}}]{taie2022observation}%
  \BibitemOpen
  \bibfield  {author} {\bibinfo {author} {\bibfnamefont {S.}~\bibnamefont
  {Taie}}, \bibinfo {author} {\bibfnamefont {E.}~\bibnamefont
  {Ibarra-Garc{\'{\i}}a-Padilla}}, \bibinfo {author} {\bibfnamefont
  {N.}~\bibnamefont {Nishizawa}}, \bibinfo {author} {\bibfnamefont
  {Y.}~\bibnamefont {Takasu}}, \bibinfo {author} {\bibfnamefont
  {Y.}~\bibnamefont {Kuno}}, \bibinfo {author} {\bibfnamefont {H.-T.}\
  \bibnamefont {Wei}}, \bibinfo {author} {\bibfnamefont {R.~T.}\ \bibnamefont
  {Scalettar}}, \bibinfo {author} {\bibfnamefont {K.~R.~A.}\ \bibnamefont
  {Hazzard}}, \ and\ \bibinfo {author} {\bibfnamefont {Y.}~\bibnamefont
  {Takahashi}},\ }\href {\doibase 10.1038/s41567-022-01725-6} {\bibfield
  {journal} {\bibinfo  {journal} {Nature Physics}\ } (\bibinfo {year} {2022}),\
  10.1038/s41567-022-01725-6}\BibitemShut {NoStop}%
\bibitem [{\citenamefont {Mukherjee}\ \emph {et~al.}(2025)\citenamefont
  {Mukherjee}, \citenamefont {Hutson},\ and\ \citenamefont
  {Hazzard}}]{mukherjee2024sun}%
  \BibitemOpen
  \bibfield  {author} {\bibinfo {author} {\bibfnamefont {B.}~\bibnamefont
  {Mukherjee}}, \bibinfo {author} {\bibfnamefont {J.~M.}\ \bibnamefont
  {Hutson}}, \ and\ \bibinfo {author} {\bibfnamefont {K.~R.~A.}\ \bibnamefont
  {Hazzard}},\ }\href {\doibase 10.1088/1367-2630/ad89f2} {\bibfield  {journal}
  {\bibinfo  {journal} {New Journal of Physics}\ }\textbf {\bibinfo {volume}
  {27}},\ \bibinfo {pages} {013013} (\bibinfo {year} {2025})}\BibitemShut
  {NoStop}%
\bibitem [{\citenamefont {Mukherjee}\ and\ \citenamefont
  {Hutson}(2025)}]{mukherjee2024sunb}%
  \BibitemOpen
  \bibfield  {author} {\bibinfo {author} {\bibfnamefont {B.}~\bibnamefont
  {Mukherjee}}\ and\ \bibinfo {author} {\bibfnamefont {J.~M.}\ \bibnamefont
  {Hutson}},\ }\href {\doibase 10.1103/PhysRevResearch.7.013099} {\bibfield
  {journal} {\bibinfo  {journal} {Phys. Rev. Res.}\ }\textbf {\bibinfo {volume}
  {7}},\ \bibinfo {pages} {013099} (\bibinfo {year} {2025})}\BibitemShut
  {NoStop}%
\bibitem [{\citenamefont {Sutherland}(1968)}]{sutherland1968further}%
  \BibitemOpen
  \bibfield  {author} {\bibinfo {author} {\bibfnamefont {B.}~\bibnamefont
  {Sutherland}},\ }\href {\doibase 10.1103/PhysRevLett.20.98} {\bibfield
  {journal} {\bibinfo  {journal} {Phys. Rev. Lett.}\ }\textbf {\bibinfo
  {volume} {20}},\ \bibinfo {pages} {98} (\bibinfo {year} {1968})}\BibitemShut
  {NoStop}%
\bibitem [{\citenamefont {Chetcuti}\ \emph
  {et~al.}(2023{\natexlab{a}})\citenamefont {Chetcuti}, \citenamefont {Polo},
  \citenamefont {Osterloh}, \citenamefont {Castorina},\ and\ \citenamefont
  {Amico}}]{chetcuti2023probe}%
  \BibitemOpen
  \bibfield  {author} {\bibinfo {author} {\bibfnamefont {W.~J.}\ \bibnamefont
  {Chetcuti}}, \bibinfo {author} {\bibfnamefont {J.}~\bibnamefont {Polo}},
  \bibinfo {author} {\bibfnamefont {A.}~\bibnamefont {Osterloh}}, \bibinfo
  {author} {\bibfnamefont {P.}~\bibnamefont {Castorina}}, \ and\ \bibinfo
  {author} {\bibfnamefont {L.}~\bibnamefont {Amico}},\ }\href {\doibase
  10.1038/s42005-023-01256-3} {\bibfield  {journal} {\bibinfo  {journal} {Comm.
  Phys.}\ }\textbf {\bibinfo {volume} {6}} (\bibinfo {year}
  {2023}{\natexlab{a}}),\ 10.1038/s42005-023-01256-3}\BibitemShut {NoStop}%
\bibitem [{\citenamefont {Peierls}(1933)}]{peierls1933zur}%
  \BibitemOpen
  \bibfield  {author} {\bibinfo {author} {\bibfnamefont {R.}~\bibnamefont
  {Peierls}},\ }\href {\doibase 10.1007/bf01342591} {\bibfield  {journal}
  {\bibinfo  {journal} {Zeitschrift für Physik}\ }\textbf {\bibinfo {volume}
  {80}},\ \bibinfo {pages} {763} (\bibinfo {year} {1933})}\BibitemShut
  {NoStop}%
\bibitem [{\citenamefont {Amico}\ \emph {et~al.}(2021)\citenamefont {Amico},
  \citenamefont {Boshier}, \citenamefont {Birkl}, \citenamefont {Minguzzi},
  \citenamefont {Miniatura}, \citenamefont {Kwek}, \citenamefont {Aghamalyan},
  \citenamefont {Ahufinger}, \citenamefont {Anderson},\ and\ \citenamefont
  {Andrei}}]{amico2021roadmap}%
  \BibitemOpen
  \bibfield  {author} {\bibinfo {author} {\bibfnamefont {L.}~\bibnamefont
  {Amico}}, \bibinfo {author} {\bibfnamefont {M.}~\bibnamefont {Boshier}},
  \bibinfo {author} {\bibfnamefont {G.}~\bibnamefont {Birkl}}, \bibinfo
  {author} {\bibfnamefont {A.}~\bibnamefont {Minguzzi}}, \bibinfo {author}
  {\bibfnamefont {C.}~\bibnamefont {Miniatura}}, \bibinfo {author}
  {\bibfnamefont {L.-C.}\ \bibnamefont {Kwek}}, \bibinfo {author}
  {\bibfnamefont {D.}~\bibnamefont {Aghamalyan}}, \bibinfo {author}
  {\bibfnamefont {V.}~\bibnamefont {Ahufinger}}, \bibinfo {author}
  {\bibfnamefont {D.}~\bibnamefont {Anderson}}, \ and\ \bibinfo {author}
  {\bibfnamefont {e.~a.}\ \bibnamefont {Andrei}},\ }\href {\doibase
  10.1116/5.0026178} {\bibfield  {journal} {\bibinfo  {journal} {AVS Q. Sci.}\
  }\textbf {\bibinfo {volume} {3}},\ \bibinfo {pages} {039201} (\bibinfo {year}
  {2021})}\BibitemShut {NoStop}%
\bibitem [{\citenamefont {Wright}\ \emph {et~al.}(2013)\citenamefont {Wright},
  \citenamefont {Blakestad}, \citenamefont {Lobb}, \citenamefont {Phillips},\
  and\ \citenamefont {Campbell}}]{wright2013driving}%
  \BibitemOpen
  \bibfield  {author} {\bibinfo {author} {\bibfnamefont {K.~C.}\ \bibnamefont
  {Wright}}, \bibinfo {author} {\bibfnamefont {R.~B.}\ \bibnamefont
  {Blakestad}}, \bibinfo {author} {\bibfnamefont {C.~J.}\ \bibnamefont {Lobb}},
  \bibinfo {author} {\bibfnamefont {W.~D.}\ \bibnamefont {Phillips}}, \ and\
  \bibinfo {author} {\bibfnamefont {G.~K.}\ \bibnamefont {Campbell}},\ }\href
  {\doibase 10.1103/PhysRevLett.110.025302} {\bibfield  {journal} {\bibinfo
  {journal} {Physical Review Letters}\ }\textbf {\bibinfo {volume} {110}},\
  \bibinfo {pages} {025302} (\bibinfo {year} {2013})}\BibitemShut {NoStop}%
\bibitem [{\citenamefont {Polo}\ \emph
  {et~al.}(2024{\natexlab{a}})\citenamefont {Polo}, \citenamefont {Chetcuti},
  \citenamefont {Haug}, \citenamefont {Minguzzi}, \citenamefont {Wright},\ and\
  \citenamefont {Amico}}]{polo2024persistent}%
  \BibitemOpen
  \bibfield  {author} {\bibinfo {author} {\bibfnamefont {J.}~\bibnamefont
  {Polo}}, \bibinfo {author} {\bibfnamefont {W.~J.}\ \bibnamefont {Chetcuti}},
  \bibinfo {author} {\bibfnamefont {T.}~\bibnamefont {Haug}}, \bibinfo {author}
  {\bibfnamefont {A.}~\bibnamefont {Minguzzi}}, \bibinfo {author}
  {\bibfnamefont {K.}~\bibnamefont {Wright}}, \ and\ \bibinfo {author}
  {\bibfnamefont {L.}~\bibnamefont {Amico}},\ }\href
  {https://arxiv.org/abs/2410.17318} {\enquote {\bibinfo {title} {Persistent
  currents in ultracold gases},}\ } (\bibinfo {year} {2024}{\natexlab{a}}),\
  \Eprint {http://arxiv.org/abs/2410.17318} {arXiv:2410.17318
  [cond-mat.quant-gas]} \BibitemShut {NoStop}%
\bibitem [{\citenamefont {Leggett}(1991)}]{leggett1991dephasing}%
  \BibitemOpen
  \bibfield  {author} {\bibinfo {author} {\bibfnamefont {A.~J.}\ \bibnamefont
  {Leggett}},\ }\enquote {\bibinfo {title} {{Dephasing and Non-Dephasing
  Collisions in Nanostructures}},}\ in\ \href@noop {} {\emph {\bibinfo
  {booktitle} {Granular Nanoelectronics}}}\ (\bibinfo  {publisher} {Springer
  US},\ \bibinfo {address} {Boston, MA},\ \bibinfo {year} {1991})\ p.\ \bibinfo
  {pages} {297}\BibitemShut {NoStop}%
\bibitem [{Note1()}]{Note1}%
  \BibitemOpen
  \bibinfo {note} {On account of this, such behaviour is different than that of
  bosons, even the ones with attractive interactions where fractionalization is
  present.}\BibitemShut {Stop}%
\bibitem [{\citenamefont {Osterloh}\ \emph {et~al.}(2023)\citenamefont
  {Osterloh}, \citenamefont {Polo}, \citenamefont {Chetcuti},\ and\
  \citenamefont {Amico}}]{osterloh2023exact}%
  \BibitemOpen
  \bibfield  {author} {\bibinfo {author} {\bibfnamefont {A.}~\bibnamefont
  {Osterloh}}, \bibinfo {author} {\bibfnamefont {J.}~\bibnamefont {Polo}},
  \bibinfo {author} {\bibfnamefont {W.~J.}\ \bibnamefont {Chetcuti}}, \ and\
  \bibinfo {author} {\bibfnamefont {L.}~\bibnamefont {Amico}},\ }\href
  {\doibase 10.21468/SciPostPhys.15.1.006} {\bibfield  {journal} {\bibinfo
  {journal} {SciPost Physics}\ }\textbf {\bibinfo {volume} {15}},\ \bibinfo
  {pages} {006} (\bibinfo {year} {2023})}\BibitemShut {NoStop}%
\bibitem [{\citenamefont {Pecci}\ \emph {et~al.}(2023)\citenamefont {Pecci},
  \citenamefont {Aupetit-Diallo}, \citenamefont {Albert}, \citenamefont
  {Vignolo},\ and\ \citenamefont {Minguzzi}}]{pecci2023persistent}%
  \BibitemOpen
  \bibfield  {author} {\bibinfo {author} {\bibfnamefont {G.}~\bibnamefont
  {Pecci}}, \bibinfo {author} {\bibfnamefont {G.}~\bibnamefont
  {Aupetit-Diallo}}, \bibinfo {author} {\bibfnamefont {M.}~\bibnamefont
  {Albert}}, \bibinfo {author} {\bibfnamefont {P.}~\bibnamefont {Vignolo}}, \
  and\ \bibinfo {author} {\bibfnamefont {A.}~\bibnamefont {Minguzzi}},\ }\href
  {\doibase 10.5802/crphys.157} {\bibfield  {journal} {\bibinfo  {journal}
  {Comptes Rendus Physique}\ }\textbf {\bibinfo {volume} {24}},\ \bibinfo
  {pages} {1} (\bibinfo {year} {2023})}\BibitemShut {NoStop}%
\bibitem [{Note2()}]{Note2}%
  \BibitemOpen
  \bibinfo {note} {Note that the SU($N$) representation can be different, which
  would generally be indicated by a different $s$. However, for some $N$ the
  different representations can share the same $s$.}\BibitemShut {Stop}%
\bibitem [{\citenamefont {Ogata}\ and\ \citenamefont
  {Shiba}(1990)}]{ogata1990bethe}%
  \BibitemOpen
  \bibfield  {author} {\bibinfo {author} {\bibfnamefont {M.}~\bibnamefont
  {Ogata}}\ and\ \bibinfo {author} {\bibfnamefont {H.}~\bibnamefont {Shiba}},\
  }\href@noop {} {\bibfield  {journal} {\bibinfo  {journal} {Physical Review
  B}\ }\textbf {\bibinfo {volume} {41}},\ \bibinfo {pages} {2326} (\bibinfo
  {year} {1990})}\BibitemShut {NoStop}%
\bibitem [{Note3()}]{Note3}%
  \BibitemOpen
  \bibinfo {note} {The barrier is the least effective versus interactions at
  $N_{p}=N$. In this case, the density at the impurity site coincides with that
  of a bosonic system of the same $N_{p}$. Such a behaviour reflects the lack
  of a meaningful Pauli exclusion principle in the system.}\BibitemShut {Stop}%
\bibitem [{\citenamefont {Litvinov}\ \emph {et~al.}(2021)\citenamefont
  {Litvinov}, \citenamefont {Bataille}, \citenamefont {Mar{\'e}chal},
  \citenamefont {Pedri}, \citenamefont {Gorceix}, \citenamefont
  {Robert-De-Saint-Vincent},\ and\ \citenamefont
  {Laburthe-Tolra}}]{litvinov2021measuring}%
  \BibitemOpen
  \bibfield  {author} {\bibinfo {author} {\bibfnamefont {A.}~\bibnamefont
  {Litvinov}}, \bibinfo {author} {\bibfnamefont {P.}~\bibnamefont {Bataille}},
  \bibinfo {author} {\bibfnamefont {E.}~\bibnamefont {Mar{\'e}chal}}, \bibinfo
  {author} {\bibfnamefont {P.}~\bibnamefont {Pedri}}, \bibinfo {author}
  {\bibfnamefont {O.}~\bibnamefont {Gorceix}}, \bibinfo {author} {\bibfnamefont
  {M.}~\bibnamefont {Robert-De-Saint-Vincent}}, \ and\ \bibinfo {author}
  {\bibfnamefont {B.}~\bibnamefont {Laburthe-Tolra}},\ }\href@noop {}
  {\bibfield  {journal} {\bibinfo  {journal} {Physical Review A}\ }\textbf
  {\bibinfo {volume} {104}},\ \bibinfo {pages} {033309} (\bibinfo {year}
  {2021})}\BibitemShut {NoStop}%
\bibitem [{\citenamefont {Chetcuti}\ \emph
  {et~al.}(2023{\natexlab{b}})\citenamefont {Chetcuti}, \citenamefont
  {Osterloh}, \citenamefont {Amico},\ and\ \citenamefont
  {Polo}}]{chetcuti2023interference}%
  \BibitemOpen
  \bibfield  {author} {\bibinfo {author} {\bibfnamefont {W.~J.}\ \bibnamefont
  {Chetcuti}}, \bibinfo {author} {\bibfnamefont {A.}~\bibnamefont {Osterloh}},
  \bibinfo {author} {\bibfnamefont {L.}~\bibnamefont {Amico}}, \ and\ \bibinfo
  {author} {\bibfnamefont {J.}~\bibnamefont {Polo}},\ }\href {\doibase
  10.21468/SciPostPhys.15.4.181} {\bibfield  {journal} {\bibinfo  {journal}
  {SciPost Phys.}\ }\textbf {\bibinfo {volume} {15}},\ \bibinfo {pages} {181}
  (\bibinfo {year} {2023}{\natexlab{b}})}\BibitemShut {NoStop}%
\bibitem [{\citenamefont {Botzung}\ and\ \citenamefont
  {Nataf}(2024)}]{botzung2024exact}%
  \BibitemOpen
  \bibfield  {author} {\bibinfo {author} {\bibfnamefont {T.}~\bibnamefont
  {Botzung}}\ and\ \bibinfo {author} {\bibfnamefont {P.}~\bibnamefont
  {Nataf}},\ }\href {\doibase 10.1103/PhysRevLett.132.153001} {\bibfield
  {journal} {\bibinfo  {journal} {Phys. Rev. Lett.}\ }\textbf {\bibinfo
  {volume} {132}},\ \bibinfo {pages} {153001} (\bibinfo {year}
  {2024})}\BibitemShut {NoStop}%
\bibitem [{\citenamefont {Weichselbaum}(2024)}]{weichselbaum2024qspace}%
  \BibitemOpen
  \bibfield  {author} {\bibinfo {author} {\bibfnamefont {A.}~\bibnamefont
  {Weichselbaum}},\ }\href {\doibase 10.21468/SciPostPhysCodeb.40} {\bibfield
  {journal} {\bibinfo  {journal} {SciPost Phys. Codebases}\ ,\ \bibinfo {pages}
  {40}} (\bibinfo {year} {2024})}\BibitemShut {NoStop}%
\bibitem [{\citenamefont {Seidel}\ and\ \citenamefont
  {Lee}(2005)}]{seidelLuther2005}%
  \BibitemOpen
  \bibfield  {author} {\bibinfo {author} {\bibfnamefont {A.}~\bibnamefont
  {Seidel}}\ and\ \bibinfo {author} {\bibfnamefont {D.-H.}\ \bibnamefont
  {Lee}},\ }\href {\doibase 10.1103/PhysRevB.71.045113} {\bibfield  {journal}
  {\bibinfo  {journal} {Phys. Rev. B}\ }\textbf {\bibinfo {volume} {71}},\
  \bibinfo {pages} {045113} (\bibinfo {year} {2005})}\BibitemShut {NoStop}%
\bibitem [{\citenamefont {Kusmartsev}(1994)}]{kusmartsev1994aharonov}%
  \BibitemOpen
  \bibfield  {author} {\bibinfo {author} {\bibfnamefont {F.~V.}\ \bibnamefont
  {Kusmartsev}},\ }\href {http://jetpletters.ru/ps/0/article_20411.shtml}
  {\bibfield  {journal} {\bibinfo  {journal} {JETP Letters}\ }\textbf {\bibinfo
  {volume} {60}},\ \bibinfo {pages} {639} (\bibinfo {year} {1994})}\BibitemShut
  {NoStop}%
\bibitem [{\citenamefont {Gogolin}\ \emph {et~al.}(2004)\citenamefont
  {Gogolin}, \citenamefont {Nersesyan},\ and\ \citenamefont
  {Tsvelik}}]{gogolin2004bosonization}%
  \BibitemOpen
  \bibfield  {author} {\bibinfo {author} {\bibfnamefont {A.~O.}\ \bibnamefont
  {Gogolin}}, \bibinfo {author} {\bibfnamefont {A.~A.}\ \bibnamefont
  {Nersesyan}}, \ and\ \bibinfo {author} {\bibfnamefont {A.~M.}\ \bibnamefont
  {Tsvelik}},\ }\href@noop {} {\emph {\bibinfo {title} {Bosonization and
  strongly correlated systems}}}\ (\bibinfo  {publisher} {Cambridge university
  press},\ \bibinfo {year} {2004})\BibitemShut {NoStop}%
\bibitem [{\citenamefont {Amaricci}\ \emph {et~al.}(2025)\citenamefont
  {Amaricci}, \citenamefont {Richaud}, \citenamefont {Capone}, \citenamefont
  {Oppong},\ and\ \citenamefont {Scazza}}]{amaricci2025engineering}%
  \BibitemOpen
  \bibfield  {author} {\bibinfo {author} {\bibfnamefont {A.}~\bibnamefont
  {Amaricci}}, \bibinfo {author} {\bibfnamefont {A.}~\bibnamefont {Richaud}},
  \bibinfo {author} {\bibfnamefont {M.}~\bibnamefont {Capone}}, \bibinfo
  {author} {\bibfnamefont {N.~D.}\ \bibnamefont {Oppong}}, \ and\ \bibinfo
  {author} {\bibfnamefont {F.}~\bibnamefont {Scazza}},\ }\href@noop {}
  {\bibfield  {journal} {\bibinfo  {journal} {arXiv preprint arXiv:2505.14630}\
  } (\bibinfo {year} {2025})}\BibitemShut {NoStop}%
\bibitem [{\citenamefont {Amico}\ \emph {et~al.}(2022)\citenamefont {Amico},
  \citenamefont {Anderson}, \citenamefont {Boshier}, \citenamefont {Brantut},
  \citenamefont {Kwek}, \citenamefont {Minguzzi},\ and\ \citenamefont {von
  Klitzing}}]{amico2022colloquium}%
  \BibitemOpen
  \bibfield  {author} {\bibinfo {author} {\bibfnamefont {L.}~\bibnamefont
  {Amico}}, \bibinfo {author} {\bibfnamefont {D.}~\bibnamefont {Anderson}},
  \bibinfo {author} {\bibfnamefont {M.}~\bibnamefont {Boshier}}, \bibinfo
  {author} {\bibfnamefont {J.-P.}\ \bibnamefont {Brantut}}, \bibinfo {author}
  {\bibfnamefont {L.-C.}\ \bibnamefont {Kwek}}, \bibinfo {author}
  {\bibfnamefont {A.}~\bibnamefont {Minguzzi}}, \ and\ \bibinfo {author}
  {\bibfnamefont {W.}~\bibnamefont {von Klitzing}},\ }\href {\doibase
  10.1103/RevModPhys.94.041001} {\bibfield  {journal} {\bibinfo  {journal}
  {Reviews of Modern Physics}\ }\textbf {\bibinfo {volume} {94}},\ \bibinfo
  {pages} {041001} (\bibinfo {year} {2022})}\BibitemShut {NoStop}%
\bibitem [{\citenamefont {Polo}\ \emph
  {et~al.}(2024{\natexlab{b}})\citenamefont {Polo}, \citenamefont {Chetcuti},
  \citenamefont {Domanti}, \citenamefont {Kitson}, \citenamefont {Osterloh},
  \citenamefont {Perciavalle}, \citenamefont {Singh},\ and\ \citenamefont
  {Amico}}]{polo2024perspective}%
  \BibitemOpen
  \bibfield  {author} {\bibinfo {author} {\bibfnamefont {J.}~\bibnamefont
  {Polo}}, \bibinfo {author} {\bibfnamefont {W.~J.}\ \bibnamefont {Chetcuti}},
  \bibinfo {author} {\bibfnamefont {E.~C.}\ \bibnamefont {Domanti}}, \bibinfo
  {author} {\bibfnamefont {P.}~\bibnamefont {Kitson}}, \bibinfo {author}
  {\bibfnamefont {A.}~\bibnamefont {Osterloh}}, \bibinfo {author}
  {\bibfnamefont {F.}~\bibnamefont {Perciavalle}}, \bibinfo {author}
  {\bibfnamefont {V.~P.}\ \bibnamefont {Singh}}, \ and\ \bibinfo {author}
  {\bibfnamefont {L.}~\bibnamefont {Amico}},\ }\href {\doibase
  10.1088/2058-9565/ad48b2} {\bibfield  {journal} {\bibinfo  {journal} {Q. Sci.
  Tech.}\ }\textbf {\bibinfo {volume} {9}},\ \bibinfo {pages} {030501}
  (\bibinfo {year} {2024}{\natexlab{b}})}\BibitemShut {NoStop}%
\bibitem [{\citenamefont {Naldesi}\ \emph {et~al.}(2022)\citenamefont
  {Naldesi}, \citenamefont {Polo}, \citenamefont {Dunjko}, \citenamefont
  {Perrin}, \citenamefont {Olshanii}, \citenamefont {Amico},\ and\
  \citenamefont {Minguzzi}}]{naldesi2022enhancing}%
  \BibitemOpen
  \bibfield  {author} {\bibinfo {author} {\bibfnamefont {P.}~\bibnamefont
  {Naldesi}}, \bibinfo {author} {\bibfnamefont {J.}~\bibnamefont {Polo}},
  \bibinfo {author} {\bibfnamefont {V.}~\bibnamefont {Dunjko}}, \bibinfo
  {author} {\bibfnamefont {H.}~\bibnamefont {Perrin}}, \bibinfo {author}
  {\bibfnamefont {M.}~\bibnamefont {Olshanii}}, \bibinfo {author}
  {\bibfnamefont {L.}~\bibnamefont {Amico}}, \ and\ \bibinfo {author}
  {\bibfnamefont {A.}~\bibnamefont {Minguzzi}},\ }\href {\doibase
  10.21468/SciPostPhys.12.4.138} {\bibfield  {journal} {\bibinfo  {journal}
  {SciPost Physics}\ }\textbf {\bibinfo {volume} {12}},\ \bibinfo {pages} {138}
  (\bibinfo {year} {2022})}\BibitemShut {NoStop}%
\end{thebibliography}%

\clearpage 


\section*{Supplementary material}

In the following, we provide supporting details of the result found in the manuscript 'Static impurity in a mesoscopic system of SU(N) fermions'.

## I. MAPPING BETWEEN THE GAUDIN-YANG-SUTHERLAND AND SU($N$) HUBBARD MODELS

Following the approach in [67], to sketch out the mapping between the lattice and continuous models we start by defining the density of $N_p$ fermions in a ring of $N_s$ sites with lattice spacing $\Delta$ as $D = N_p/(\Delta N_s)$. Subsequently, the filling fraction $\nu = N_p/N_s$ can be reformulated in terms of the lattice spacing $\nu = D\Delta$. As a result, in the continuous limit of vanishing lattice spacing, $\Delta \to 0$, and for finite particle density, the corresponding filling fraction must also be small. In order for the anti-commutation relations to be preserved in the continuous limit, one needs to re-scale the fermionic operators by the lattice spacing in the following manner

$$c_{j,\alpha}^\dagger = \sqrt{\Delta}\Psi_\alpha^\dagger(x_j) \qquad n_{j,\alpha} = \Delta\Psi_\alpha^\dagger(x_j)\Psi_\alpha(x_j) \qquad \text{where } x_j = j\Delta \tag{2}$$

with $\Psi_\alpha^\dagger(x_j)$ [$\Psi_\alpha(x_j)$] being the field operators that create (annihilate) a fermion with effective spin projection $\alpha$ at position $x_j$, obeying the standard anti-commutation relations $\{\Psi_\alpha(x), \Psi_\beta^\dagger(y)\} = \delta_{\alpha,\beta}\delta(x-y)$ and $\{\Psi_\alpha^\dagger(x), \Psi_\beta^\dagger(y)\} = 0$. Plugging the re-scaled fermionic operators of Eq. (2) in the SU($N$) Hubbard model, we have that the Hamiltonian is mapped onto the Fermi gas quantum field theory $\mathcal{H}_{FG}$ via $\mathcal{H}_{SU(N)} = t\Delta^2 \mathcal{H}_{FG} - 2N_p$ where

$$\mathcal{H}_{FG} = \int \left[ -(\partial_x \Psi_\alpha^\dagger)(\partial_x \Psi_\alpha) + 2c\sum_{\alpha<\beta}^N \Psi_\alpha^\dagger \Psi_\beta^\dagger \Psi_\beta \Psi_\alpha \right] \mathrm{d}x, \tag{3}$$

where $\alpha$ and $\beta$ correspond to the different colours and we defined the interaction strength as $c = \frac{U}{4\Delta t}$ with $t$ and $U$ being the Hubbard model's hopping and interaction strengths respectively. The eigenstates of $\mathcal{H}_{FG}$ are of the given form

$$|\psi(\lambda)\rangle = \sum_{\alpha_1\ldots\alpha_{N_p}}^N \int \chi(\mathbf{x}|\lambda) \Psi_{\alpha_1}^\dagger(x_1)\ldots\Psi_{\alpha_{N_p}}^\dagger(x_{N_p})|0\rangle \mathrm{d}\mathbf{x}, \tag{4}$$

with $\chi(\mathbf{x}|\lambda)$ being eigenfunctions of the Gaudin-Yang-Sutherland model that reads

$$\mathcal{H}_{GYS} = -\frac{\hbar^2}{2m}\sum_{\alpha=1}^N \sum_{j=1}^{N_\alpha} \frac{\partial^2}{\partial x_{j,\alpha}^2} + 2c\sum_{\alpha<\beta}^N \sum_{j,k}^{N_p} \delta(x_{j,\alpha} - x_{k,\beta}), \tag{5}$$

where $m$ denotes the mass of the particles and $N_\alpha$ is the number of particles per colour.

## II. BETHE ANSATZ EQUATIONS OF THE GAUDIN-YANG-SUTHERLAND MODEL WITH TWISTED BOUNDARY CONDITIONS

In the presence of an effective magnetic flux $\phi$, modifies the momentum of the fermions due to its coupling with the vector potential. Consequently, the Gaudin-Yang-Sutherland model of Eq. (5) can be re-written as

$$\mathcal{H}_{GYS} = -\frac{\hbar^2}{2m}\sum_{\alpha=1}^N \left[\sum_{j=1}^{N_\alpha}\left(\frac{\partial}{\partial x_{j,\alpha}} - \frac{2\pi}{L}\frac{\phi}{\phi_0}\right)^2 + 2c\sum_{\alpha<\beta}^N \sum_{j,k}^{N_p} \delta(x_{k,\alpha} - x_{j,\beta}), \tag{6}$$

with $\phi_0$ being the elementary flux quantum. The model is Bethe Ansatz integrable [47] through the following equations

$$e^{i(k_j L - 2\pi\phi)} = \prod_{\alpha=1}^{M_1} \frac{k_j - \Lambda_\alpha^{(1)} + ic}{k_j - \Lambda_\alpha^{(1)} - ic}, \quad j = 1,\ldots,N_p \tag{7}$$

$$\prod_{\substack{\beta=1 \\ \beta\neq\alpha}}^{M_r} \frac{\Lambda_\alpha^{(r)} - \Lambda_\beta^{(r)} + 2ic}{\Lambda_\alpha^{(r)} - \Lambda_\beta^{(r)} - 2ic} = \prod_{\beta=1}^{M_{r-1}} \frac{\Lambda_\alpha^{(r)} - \Lambda_\beta^{(r-1)} + ic}{\Lambda_\alpha^{(r)} - \Lambda_\beta^{(r-1)} - ic} \cdot \prod_{\beta=1}^{M_{r+1}} \frac{\Lambda_\alpha^{(r)} - \Lambda_\beta^{(r+1)} + ic}{\Lambda_\alpha^{(r)} - \Lambda_\beta^{(r+1)} - ic}, \quad \alpha = 1,\ldots,M_r \tag{8}$$

for $r = 1, \ldots, N - 1$ where $M_0 = N_p$, $M_N = 0$ and $\lambda_\beta^{(0)} = k_\beta$. $M_r$ corresponds to the number of spin rapidities per colour with $k_j$ and $\Lambda_\alpha^{(r)}$ being the charge and spin momenta respectively. The energy of the system is given by $E = \sum_j^{N_p} k_j^2$. Let us take the SU(3) case as an example, wherein we have three non-linear equations,

$$e^{i(k_j L - \phi)} = \prod_{\alpha=1}^{M_1} \frac{k_j - \Lambda_\alpha + ic}{k_j - \Lambda_\alpha - ic} \quad j = 1, \ldots, N_p, \tag{9}$$

$$\prod_{\beta \neq \alpha}^{M_1} \frac{\Lambda_\alpha - \Lambda_\beta + 2ic}{\Lambda_\alpha - \Lambda_\beta - 2ic} = \prod_{j=1}^{N_p} \frac{\Lambda_\alpha - k_j + ic}{\Lambda_\alpha - k_j - ic} \prod_{a=1}^{M_2} \frac{\Lambda_\alpha - \lambda_a + ic}{\Lambda_\alpha - \lambda_a - ic} \quad \alpha = 1, \ldots, M_1, \tag{10}$$

$$\prod_{b \neq a}^{M_2} \frac{\lambda_a - \lambda_b + 2ic}{\lambda_a - \lambda_b - 2ic} = \prod_{\alpha=1}^{M_1} \frac{\lambda_a - \Lambda_\alpha + ic}{\lambda_a - \Lambda_\alpha - ic} \quad a = 1, \ldots, M_2. \tag{11}$$

Note that for the sake of convenience, we changed $\Lambda_\alpha^{(1)}$ and $\Lambda_\alpha^{(2)}$ to $\Lambda_\alpha$ and $\lambda_a$. The first equation pertains to the charge quasimomenta with the other two accounting for the spin degrees of freedom, that emerge from the nested nature of the Bethe Ansatz. In the limit $\frac{U}{N_p} \to \infty$, we can neglect the terms $k_j/U$ as they are much smaller in magnitude compared to the spin momenta and will tend to zero. As a result, the Bethe Ansatz equations simplify and are re-cast into the form

$$e^{i(k_j L - \phi)} = \prod_{\alpha=1}^{M_1} \frac{\Lambda_\alpha - ic}{\Lambda_\alpha + ic} \quad j = 1, \ldots, N_p, \tag{12}$$

$$\prod_{\beta \neq \alpha}^{M_1} \frac{\Lambda_\alpha - \Lambda_\beta + 2ic}{\Lambda_\alpha - \Lambda_\beta - 2ic} = \left[\frac{\Lambda_\alpha + ic}{\Lambda_\alpha - ic}\right]^{N_p} \prod_{a=1}^{M_2} \frac{\Lambda_\alpha - \lambda_a + ic}{\Lambda_\alpha - \lambda_a - ic} \quad \alpha = 1, \ldots, M_1, \tag{13}$$

$$\prod_{b \neq a}^{M_2} \frac{\lambda_a - \lambda_b + 2ic}{\lambda_a - \lambda_b - 2ic} = \prod_{\alpha=1}^{M_1} \frac{\lambda_a - \Lambda_\alpha + ic}{\lambda_a - \Lambda_\alpha - ic} \quad a = 1, \ldots, M_2. \tag{14}$$

Essentially, there is a separation of the charge and spin degrees of freedom that decouples the rapidities in the Bethe Ansatz equations. Consequently, in this limit, the states Gaudin-Yang-Sutherland model can be written as a composition of the spinless model and SU(3) Heisenberg Hamiltonian [57, 60]. Indeed, Eqs. (13) and (14) correspond to the Bethe equations of the isotropic SU(3) Heisenberg chain barring a normalization constant [68].

Taking the logarithm of Eqs. (12) through (14) and utilizing the relation $2i \arctan x = \pm\pi + \ln \frac{x-i}{x+i}$, the quasimomenta $k_j$ can be expressed [37] as

$$k_j L = 2\pi \left[ I_j + \frac{1}{N_p}\left(\sum_{\alpha=1}^{M_1} J_\alpha + \sum_{a=1}^{M_2} L_a\right) + \phi \right], \tag{15}$$

with $I_j$ and $\{J_\alpha, L_a\}$ being the quantum numbers parameterizing the Bethe equations, the first being associated with the charge quasimomenta and the others for the spin rapidities. Therefore, the energy of the system reads

$$E(\phi) = \left(\frac{2\pi}{L}\right)^2 \left[ \sum_j^{N_p} I_j^2 + 2\sum_j^{N_p} I_j \left(\frac{X}{N_p} + \phi\right) + N_p \left(\frac{X}{N_p}\right)^2 + N_p \left(2\phi \frac{X}{N_p} + \phi^2\right) \right], \tag{16}$$

and consequently, the corresponding persistent current defined at zero temperature as $I(\phi) = -\partial E/\partial \phi$ acquires the following form in the regime of $U \to \infty$

$$I(\phi) = -\left(\frac{2\pi}{L}\right)^2 \left[ 2\sum_j^{N_p} I_j + N_p \left(2\frac{X}{N_p} + 2\phi\right) \right], \tag{17}$$

where $X = \left(\sum_{\alpha=1}^{M_1} J_\alpha + \sum_{a=1}^{M_2} L_a\right)$. For the general case of SU($N$) fermions, the expression for the persistent current is the same keeping in mind that $X = \sum_l^{N-1} \sum_{\alpha_l}^{M_l} J_{\beta_l}$ to account for the quantum numbers of the $N-1$ spin rapidities. Although the SU($N$) Hubbard model is not integrable, in the limit of infinite repulsion it is quite close to being to at least for the low-lying spectrum. Thus, by using the Bethe equations of the Lai-Sutherland model [69] corresponding to the SU($N$) Hubbard model with commensurate filling fractions, acquired by substituting $\lambda_\beta^{(0)} = \sin k_\beta$ and $k_j$ with $\sin k_j$ in Eq. (7) due to lattice regularization [68]. Following the same logic applied to the continuous regime, the quasimomenta $k_j$ turn out to be of the same form as in Eq. (15), which when plugged into the energy $E = -2\sum_j \cos k_j$ and subsequently differentiating with the flux gives us

$$I(\phi) = -\frac{2\pi}{N_s} E_m \sin\left[\frac{2\pi}{N_s}\left(D + \frac{X}{N_p} + \phi\right)\right], \tag{18}$$

where $E_m = 2\sin\left(\frac{N_p \pi}{L}\right)/\sin\left(\frac{\pi}{L}\right)$ and $D = \frac{I_{max}+I_{min}}{2}$ [37, 70][71]. Naturally, for a fixed number of particles and increasing lattice sites in the system, Eq. (18) tends to Eq. (17).

## III. ANALYSIS OF THE DENSITY AT THE SITE OF THE LOCALIZED IMPURITY

In this section, we assess the influence of the impurity, whether it's strengthened or weakened, by looking at the density on the site where it is residing, denoted by $n_{\text{imp}}$.

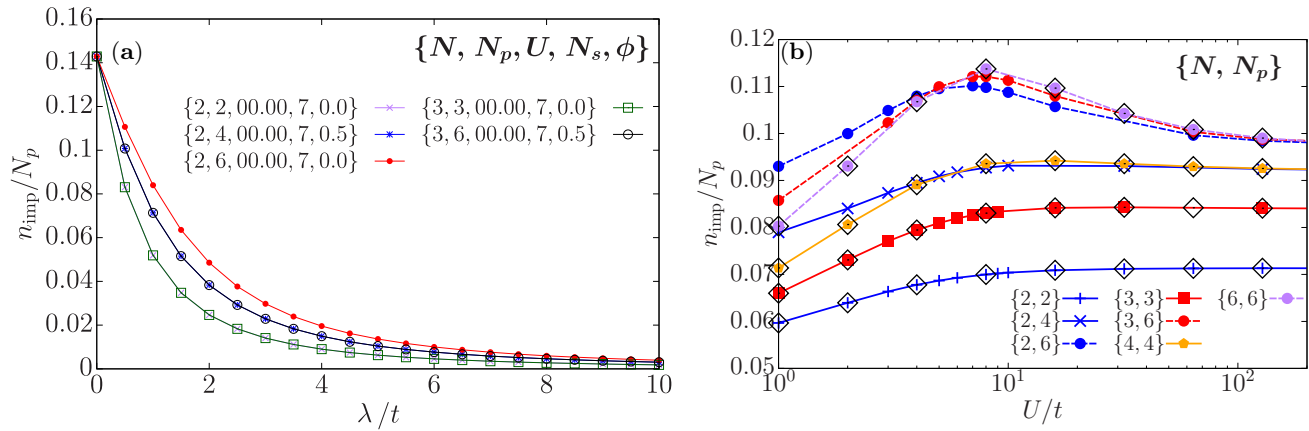

FIG. 4. *Density at the site of the impurity dependence on the interplay between barrier strength and interaction.* **(a)** Figure depicts the density at the barrier $n_{\text{imp}}$ normalized by the number of particles $N_p$ in a ring composed of $N_s$ sites for non-interacting fermions as a function of the barrier strength $\lambda$ in the absence of flux $\phi$ and different number of components $N$. **(b)** $n_{\text{imp}}$ for fixed $\lambda/t = 1$ as a function of $U/t$ for different $N_p$ and $N$ at $\phi = 0$. Grey triangles correspond to single-component bosons with the same $N_p$. Results obtained with exact diagonalization of the SU($N$) Hubbard and Bose-Hubbard model. *Note: The cases with even $N_p/N$ are recorded at $\phi/\phi_0 = 0.5$ due to a parity effect.*

For free fermions at zero flux and zero barrier strength $\lambda$, the normalized density $n_{\text{imp}}$ for different cases of number of components $N$ and particles $N_p$ is uniform throughout with a density of $1/N_s$. On increasing $\lambda$, $n_{\text{imp}}$ displays a monotonic behaviour as the role of the static impurity is to halt the flow of particles. Several interesting trends appear in Fig. 4**(a)**. Firstly, cases with the same $N_p/N$ coincide with one another. Secondly, the barrier exerts its influence more on decreasing $N_p/N$, i.e. going to larger $N$. At zero interactions, one can picture the system with $N$ components as $N$ rings that are essentially free from one another. As a result, each ring will pay an energy penalty of $\lambda$ amounting to a total contribution of $N\lambda$, enticing the particles to avoid the barrier more. Therefore, for a fixed particle number, the barrier will be more effective upon increasing $N$. An alternative way of seeing this is that the number of particles that can reside at the impurity site is limited by the Pauli exclusion principle thereby limiting the barrier's impact.

Switching on interactions, the density profile at the impurity site is monotonic for systems where $N_s \gg N_p$ – Fig. 4**(b)**. Going to weak interactions, the barrier is screened due to the repulsion between the particles. The $n_{\text{imp}}$ at equal $N_p/N$ split and for a given $N_s$ re-arrange themselves with increasing $N_p$ displaying a further ordering dictated by $N$, still observing the trend of the impurity being weakened for larger $N_p/N$. However, we remark that since the interactions are effectively enhanced with $N$, the rate at which the barrier is suppressed increases with $N$ and for fixed

number of components is governed by $N_p$ –Fig. 5(c). Indeed, the rate grows approaching $N_p/N = 1$ corresponding to a "bosonic" limit. Consequently, there are crossings between the curves of $n_{\rm imp}$ with the most notable being for fixed $N_p$ and various $N$ reversing the trend of $N_p/N$ present in the non-interacting regime –Fig. 4(b). Pushing interactions even further as $U \to \infty$, the distinction between $N$ is disregarded due to forming a collective state comprised of a stiff particle arrangement arising from the spin correlations.

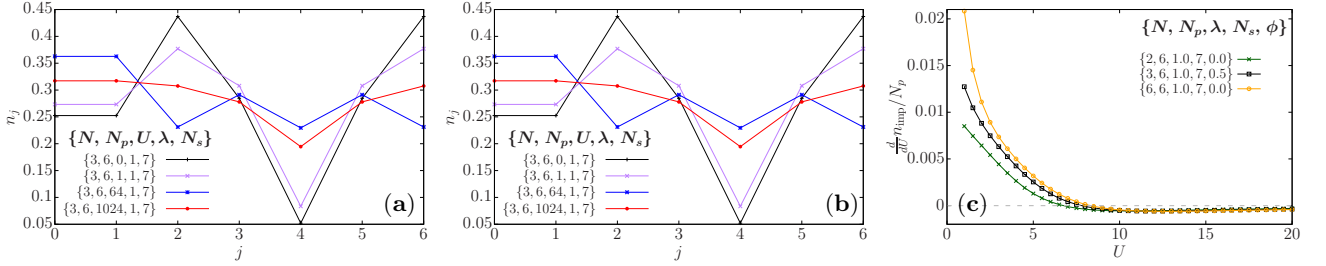

FIG. 5. *Density at each site of the ring for various interactions.* Figure showcases the density $n_j$ at each site of the ring of $N_s$ sites with fixed barrier strength $\lambda/t$ and various interactions $U/t$ in the absence of flux for $N_p = 6$ particles with different number of components (a) $N = 2$ and (b) $N = 3$. The global minimum is located on the site of the impurity at $N_s = 4$. (c) Rate of change of density with interaction denoted by $\frac{d}{dU} n_{\rm imp}$. Results obtained with exact diagonalization of the SU($N$) Hubbard model.

In the cases where $N_s$ is comparable to $N_p$, $n_{\rm imp}$ displays a non-monotonous behaviour originating as a finite size effect –Fig. 5(c). In order for the fermions to minimize the increased repulsion between them, they occupy the impurity site where the barrier essentially acts as a shield, since in this regime the particle occupation per site becomes restricted. Once again, the rate at which the barrier is augmented is faster with $N$ due to the enhanced effective repulsion. For larger system sizes, the non-monotonicity is no longer there as depicted in Fig. 6.

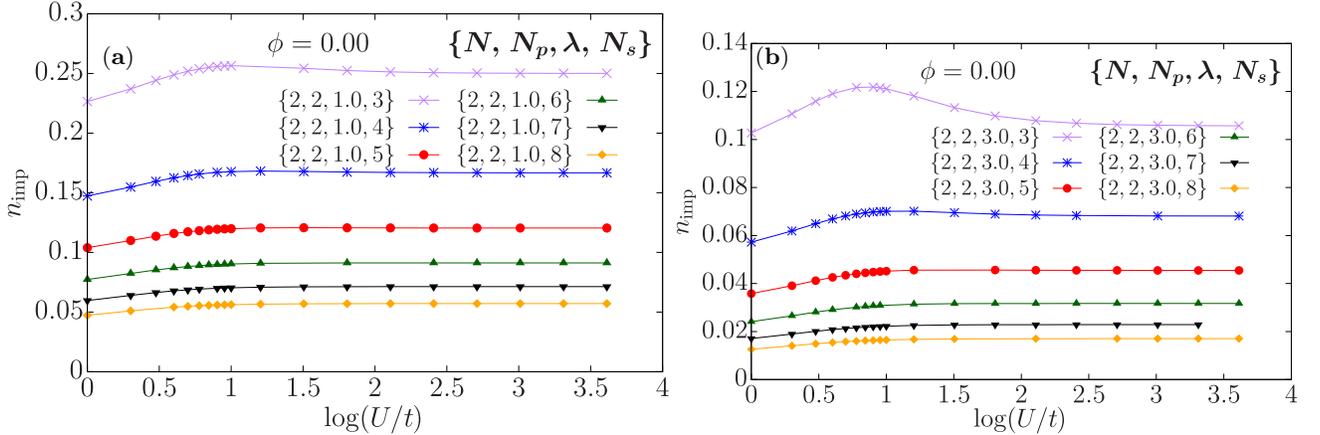

FIG. 6. *Density at the site of the impurity as a function of the number of sites and interaction.* Figure depicts the density at the barrier $n_{\rm imp}$ as a function of the interaction $U/t$ with increasing lattice sites $N_s$ for a fixed barrier strength $\lambda/t = 1$ and $\lambda/t = 3$ in panels (a) and (b) respectively in the absence of flux $\phi$. Results obtained with exact diagonalization of the SU($N$) Hubbard.

A very interesting feature emerges in the density when subjected to an effective magnetic flux $\phi$ in the presence of a barrier. Specifically, the profile of $n_{\rm imp}$ becomes periodic with $\phi$ reflecting the period of the energy and persistent current, which is given by the bare flux quantum $\phi_0$ in the free fermion regime and the fractional flux quantum $\phi_0/N_p$ observed at strong interactions –Fig. 7. As the current in the system grows larger, corresponding to a bigger particle flow, the impact of the barrier is enhanced reaching a maximum at the $\phi$, where the energy level crossing occurs, in the absence of the impurity. Subsequently, a reduction in current is portrayed by a weakened impact of the barrier. Through this, we observe that the barrier captures the periodic nature of the current and provides an alternative route to monitor fractionalization.

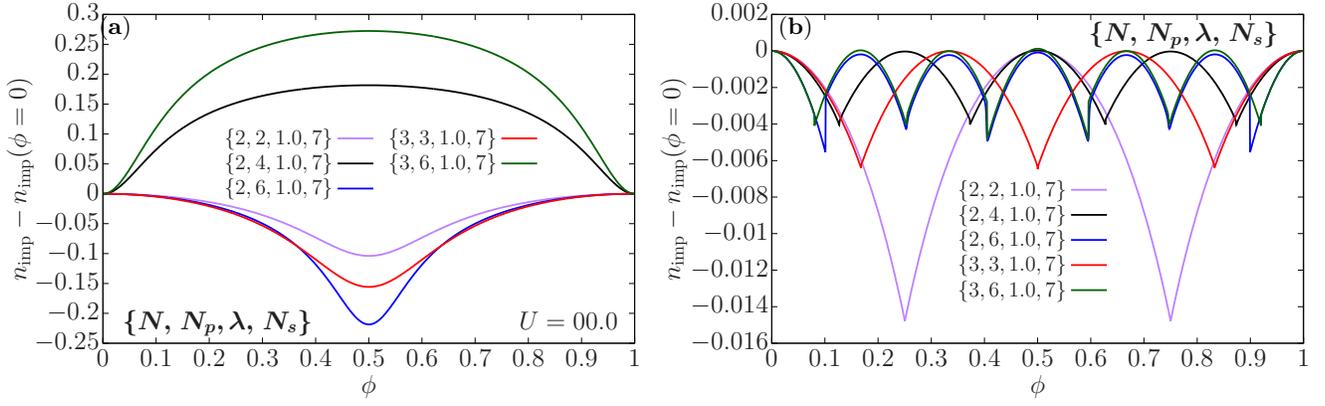

FIG. 7. *Density at the site of the impurity as a function of the flux.* Depiction of the density at the barrier site $n_{\text{imp}}$ for fixed barrier strength $\lambda/t = 1$ as a function of the flux $\phi$ for different number of particles $N_p$ and components $N$ at interactions **(a)** $U/t = 0$ and **(b)** $U/t = 1024$. Results obtained with exact diagonalization of the SU($N$) Hubbard.

## IV. SU($N$) FERMIONIC PERSISTENT CURRENTS IN A SYSTEM WITH A LOCALIZED IMPURITY

The second probe utilized in our analysis is the persistent current $I(\phi)$. Firstly, we start by remarking that in the non-interacting regime the periodicity of the current as a function of the flux, with a period fixed by $\phi_0$ originates from energy level crossings between parabolas of different angular momenta. This occurs to counteract the flux and minimize the energy of the system. In the case of fermions, these degeneracy points depend on their population parity occurring at $\phi = (2n+1)/2$ or $\phi = n$ when the number of particles $N_p = (2m+1)/N$ and $N_p = (2m)N$ prompting a diamagnetic and paramagnetic response, respectively. In the presence of a barrier and zero interactions, the opening of the spectral gap at these degeneracy points smears the characteristic saw-tooth shape of the current eventually turning into a sinusoid shape as the barrier strength $\lambda$ becomes sufficiently large –Fig. 8**(a)**. Additionally, the magnitude of the current decreases with increasing $\lambda$ as is expected.

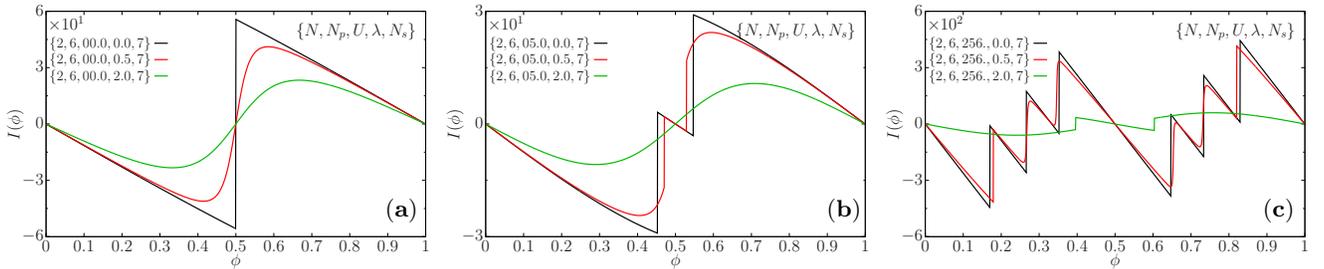

FIG. 8. *Persistent current against flux for a fixed interaction.* Figure shows the persistent current $I(\phi)$ as a function of the flux $\phi$ for different barrier strengths $\lambda$ at fixed interactions **(a)** $U/t = 0$, **(b)** $U/t = 1$ and **(c)** $U/t = 10$. Results obtained with exact diagonalization of the SU($N$) Hubbard model for $N_p = 2$ particles with SU(2) symmetry in a ring of $N_s = 7$ sites.

In the absence of a barrier, strongly interacting fermionic matter-wave currents undergo fractionalization stemming from the spin correlations in the system resulting in them having a reduced period, which depends on whether the system has repulsive or attractive interactions [38]. For the repulsive case under consideration here, $N_p$ piece-wise parabolic segments appear in the energy spectrum per flux quantum reflecting the fractional values of the angular momentum per particle. On account of this phenomenon, there is an intriguing interplay between the interaction and the impurity. Specifically, the profile of the current becomes hybridized showcasing both smoothening and fractionalization –Figs. 8 and 9. Such a behaviour is similar to the profile of matter-wave currents observed in the extended SU($N$) Hubbard model in the commensurate regime of one particle per site [72]. Eventually as $U \to \infty$, the effect of the barrier is significantly diminished and the current exhibits the characteristic $1/N_p$ periodicity. However, we note that while the value of the persistent current saturates to a value close to that obtained through Bethe Ansatz in the absence of a barrier, it does not coincide exactly.

To properly visualize the interplay between interaction and impurity, we resort to plotting the maximum current

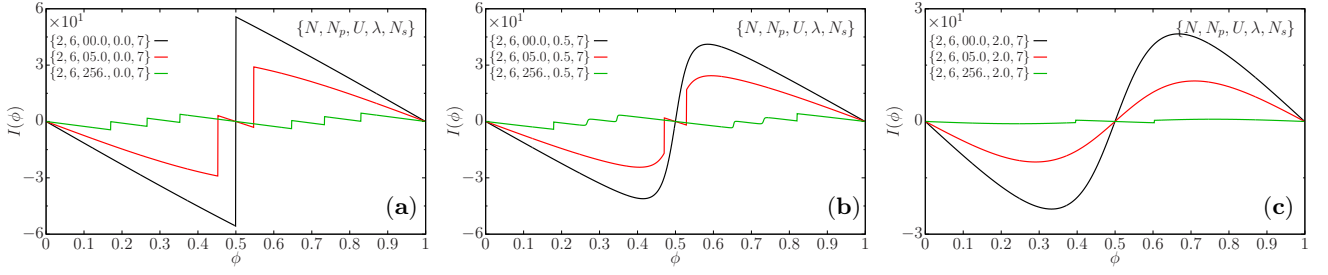

FIG. 9. *Persistent current against flux for a fixed barrier*. Figure illustrates the persistent current $I(\phi)$ as a function of the flux $\phi$ for different interactions $U/t$ at fixed barrier strength **(a)** $\lambda/t = 0$, **(b)** $\lambda/t = 1$ and **(c)** $\lambda/t = 10$. Results obtained with exact diagonalization of the SU($N$) Hubbard model for $N_p = 2$ particles with SU(2) symmetry in a ring of $N_s = 7$ sites.

amplitude $I_{\max}$ that displays a non-monotonic behaviour –Fig. 10**(a)** and **(b)**. Firstly, we point out that $I_{\max}$ at zero barrier and interactions is larger on increasing $N_p$ and for fixed particle number is greater on going to $N_p/N = 1$ [51]. At weak $U$ and a given $\lambda$, the barrier is suppressed for the same reason presented in Sec. III at rate that firstly depends on $N$ and then on $N_p$. From Fig. 10**(c)**, it is clear that the barrier is screened out faster for larger $N$ due to the increased effective repulsion up to $NU$ arising from the loosened Pauli exclusion principle. We emphasize that the fractionalization does not play a role as the gaps that open up inhibit energy level crossings, which will now occur at a larger value of interaction compared to the one in the absence of a barrier. Indeed, this is evidenced by the currents for $N_p/N = 1$ and that of bosons with same $N_p$ coinciding in this regime –Fig. 10**(a)**. For single-component bosons subjected to repulsive interactions there is no such phenomenon of fractionalization.

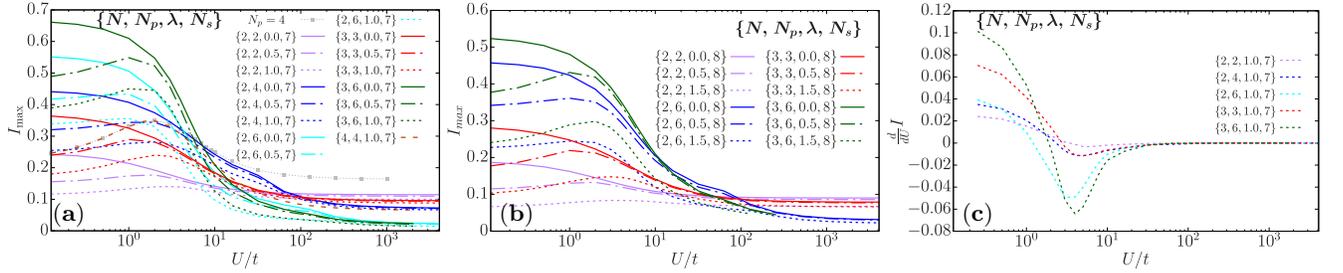

FIG. 10. *Persistent current behavior dependence on the interplay between interaction and impurity*. Persistent current amplitude $I_{\max}$ as a function of the interaction $U/t$ for different barrier strengths $\lambda/t$, number of particles $N_p$ and number of components $N$ in a ring of **(a)** $N_s = 7$ and **(b)** $N_s = 8$ sites. **(c)** Rate of change of the current $\frac{d}{dU}I_{\max}$ with interactions for fixed barrier strength. Results obtained with exact diagonalization of the SU($N$) Hubbard.

At the bump corresponding to intermediate $U$, we observe that $I_{\max}$ of the bosons and corresponding fermionic case with $N_p/N = 1$ start to separate indicating with the amplitude of the latter system being more reduced. Such behaviour indicates that the fractionalization is now playing a key role with the impact of the barrier being severely diminished. There are three indicators in this regard: **(i)** for $\lambda = 0$ fractionalization decreases the current magnitude –Fig. 10**(a)** and **(b)**; **(ii)** cases with same $N_p$ but different $N$ overlap reflecting the $1/N_p$ periodicity as $U \to \infty$; **(iii)** values of the current approach those calculated from Bethe Ansatz from Eq. (18). Lastly, the trend observed in the weakly interacting regime of $I_{\max}$ growing with $N_p$ is now reversed.

In addition to the phenomenon of fractionalization, the current magnitude is decreased due to strong quantum phase fluctuations. Seeing as for strong interactions the system is in a collective state of a stiff particle arrangement, there are no density fluctuations, hence prompting ones in the phase. Such is the case for repulsive bosons [35]. To understand this better, we can consider the quantum phase model $\mathcal{H}_{QPM}$, which the Bose-Hubbard Hamiltonian maps to in the limit $\bar{n} = N_p/N_s \gg 1$ [73]. By writing the annihilation operator in the Bose-Hubbard model as $a_j \sim \sqrt{\bar{n}} e^{i\varphi_j}$, we have that

$$\mathcal{H}_{QPM} = -2J \sum_{\langle i,j \rangle} \cos(\varphi_i - \varphi_j) + \frac{U}{2} \sum_i \delta n_i^2, \tag{19}$$

where $\varphi_i$ and $\delta n_j$ are conjugate operators for the phase phase and density fluctuations respectively with $[\varphi_i, \delta n_j] = i\hbar \delta_{ij}$. The parameters $J = t\bar{n}$ and $\delta n_j = \bar{n} - n_j$ correspond to the Josephson coupling and density fluctuations. From

here, it becomes clear that as the interactions increase the number fluctuations on the site become smaller compared to those of the phase as otherwise the energy would increase. Furthermore, we note that quantum phase model is used to describe the physics of the Josephson junction, which is essentially the tunneling of particles through a barrier.

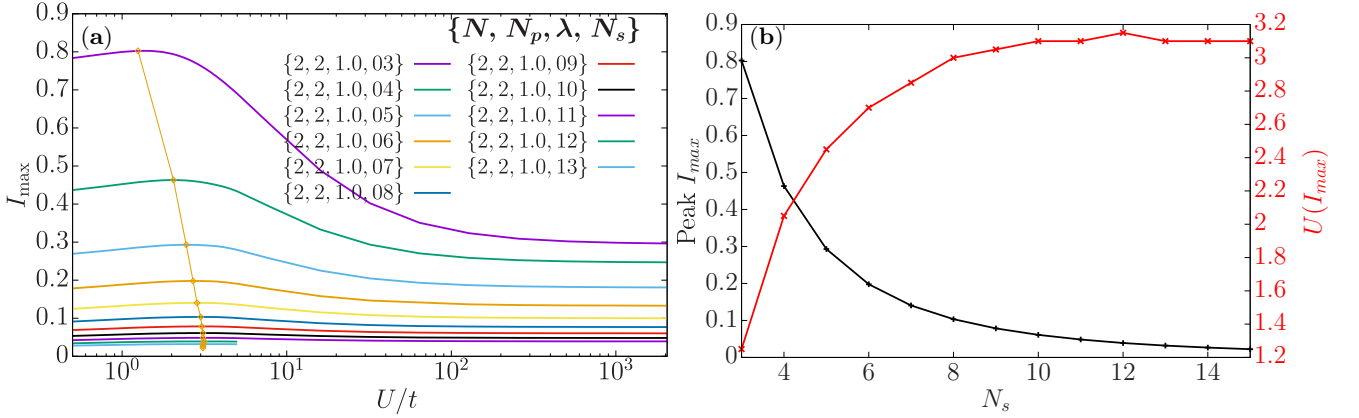

FIG. 11. *Optimal persistent current dependence on increasing system size.* **(a)** Figure illustrates the changes in the peak of the current amplitude $I_{\max}$ on increasing the number of sites in the ring $N_s$ as a function of the interaction $U/t$ for $N_p = 2$ particles with SU(2) symmetry and fixed barrier strength $\lambda/t$. orange line acts as a guide to the eye. **(b)** Change in the magnitude peak of $I_{\max}$ (black) and the interaction at which it is obtained (red) as a function of $N_s$. Results obtained with exact diagonalization of the SU($N$) Hubbard.

To conclude, we remark that the optimal current as a function of $U$ and $\lambda$, corresponding to the peak of $I_{\max}$, decreases as a function of the number of sites $N_s$. Naturally, this is to be expected as the current is a mesoscopic effect with a $1/N_s$ dependence as shown by Eq. 18. On the other hand, the interaction at which this optimal current is obtained, grows with $N_s$ until it saturates at large $N_s$. This saturation can be understood from the density where the barrier is suppressed with interaction monotonously when $N_s$ is larger than $N_p$.

## V. ANALYSIS OF THE SPECTRAL GAP

Typically, the introduction of a static impurity opens a gap $\Delta$ at all degeneracy points in the spectrum. Such is the case for single-component bosons with repulsive and attractive interactions, as well as for spinless fermions. When considering SU($N$) fermions, a condition is imposed for a gap to appear: the effective total spin, computed through the quadratic Casimir, corresponding to the intersecting parabolas needs to be the same to have the avoided level crossings.

The total quadratic Casimir operator is defined as

$$C_{1,\text{tot}} = \left(\sum_i \vec{S}_i\right)^2 = \sum_i \vec{S}_i^2 + 2\sum_{i \neq j} \vec{S}_i \cdot \vec{S}_j, \tag{20}$$

where $\vec{S}_i := \frac{1}{2} c^\dagger_{i,\alpha} \vec{\sigma}_{i;\alpha,\beta} c_{i;\beta}$ using the Einstein sum convention and $\vec{\sigma}_i$ are the SU($N$) extensions of the Pauli matrices. On a given site $i$, we have that

$$\vec{S}_i^2 = \frac{N-1}{2N} \sum_\alpha^N n_{i,\alpha}\left(1 - \sum_\beta^N n_{i,\beta}\right), \tag{21}$$

which turns out to be a function of number operators. In our setting, the static impurity is provided solely through the local density $\mathcal{H}_{\text{imp}} = \sum_\alpha^N \lambda_\alpha n_{0,\alpha}$ where $\lambda_\alpha$ is the impurity strength for species $\alpha$ and $n_0$ is the number operator denoting that the impurity is located on site 0. It is clear that impurity Hamiltonian commutes with $\vec{S}_i^2$ since both of them contain just number operators. We are left with calculating $[\sum_{i \neq j} \vec{S}_i \cdot \vec{S}_j, \sum_\alpha^N \lambda_\alpha n_{0,\alpha}]$, which results to be

$$\left[\sum_{i \neq j} \vec{S}_i \cdot \vec{S}_j, \sum_\alpha^N \lambda_\alpha n_{0,\alpha}\right] = \sum_\epsilon^{N^2-1} \sum_{j \neq 0}^{N_p} \sum_{\alpha,\beta,\gamma}^N \sigma^\epsilon_{0;\alpha,\beta} \sigma^\epsilon_{j;\beta,\gamma} c^\dagger_{j,\beta} c_{j,\gamma} c^\dagger_{0,\alpha} c_{0;\beta} (\lambda_\beta - \lambda_\alpha). \tag{22}$$

Hence, the commutator is equal to zero *iff* the impurity's strength is independent of the species. Otherwise, if it is not, the SU($N$) symmetry is broken explicitly. Considering the conditions of Ref. [57], where an equal number of particles per species $N_\alpha$ is chosen, the alculation of the quadratic Casimir simplifies to

$$C_{1,\text{tot}} = \sum_\alpha^N S_{\text{tot}}^\alpha S_{\text{tot}}^\alpha = \sum_{\alpha=1}^{N(N-1)} \left(S_{xy;\text{tot}}^\alpha\right)^2 + \sum_{\gamma=1}^{N-1} \left(S_{C;\text{tot}}^\gamma\right)^2, \tag{23}$$

where $S_{C;\text{tot}}^\gamma$ symbolizes the $N-1$ diagonal Cartan elements of the generators and $S_{xy;\text{tot}}^\alpha$ the remaining part. Since $S_{C;\text{tot}}^\gamma$ gives zero weight for the states that we consider, the quadratic Casimir reduces to

$$C_{1,\text{tot}} = \sum_{\alpha=1}^{N(N-1)} \left(S_{xy;\text{tot}}^\alpha\right)^2. \tag{24}$$

Since the barrier commutes with the quadratic Casimir operator in the cases under our consideration of species-independent $\lambda$, then it can only couple states *iff* the corresponding eigenstates share the same Casimir value[74]. Whilst at zero interaction, all energy level crossings in the ground-state have the same Casimir value, which is marked by the smoothening of the current, the same cannot be said on going to stronger $U$. On account of fractionalization, mediated by various energy level crossings from the spin correlations, the $N_p$ piece-wise parabolas are characterized by different Casimir values especially on going to larger SU($N$) resulting in selective gap openings –Fig. 12.

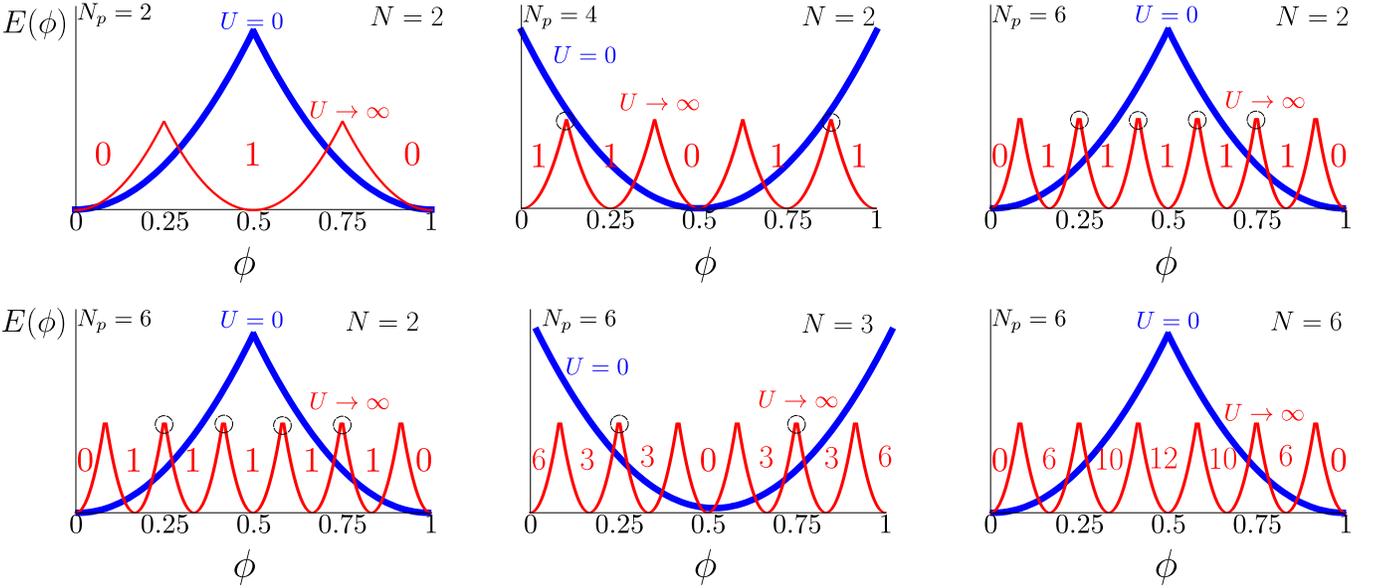

FIG. 12. schematic figure for the energy landscape $E(\phi)$ as a function of the effective magnetic flux at interactions $U/t = 0$ and $U/t \to \infty$ for $N_p$ particles and $N$ components. The numbers in each parabolas correspond to the Casimir value associated to it with the black dotted circles indicating where a gap opens up.

Fixing the barrier and going to weak interactions, we have that the gap opening depends on the parity and the number of components whereby the system can behave in two ways. For *diamagnetic parity and odd $N$*, the energy level crossings due to fractionalization intersect each other at $\phi = 0.5$ [51], such that the adjacent parabola have the same Casimir values opening a spectral gap. It turns out, that such behaviour is an exception as in all other cases[75] the spectral gap remains closed at all intersection points, at difference with the bosonic counterpart. Consequently, the spectral gap observed in the free-particle regime is now pushed towards highly excited states as there are no adjacent parabolas with the same Casimir representation. Nonetheless, we point out that this gap still produces some deformation of the energy parabolas and a weak smoothening of the current.

Ramping up to stronger interactions, more parabolas descend to the ground-state due to fractionalization to given $N_p$ energy level crossings. Consequently, there are more gaps that can be opened if the condition of having the same Casimir value is met. For a given number of particles and different component number, we find that the number of gap openings and their subsequent flux dependence, i.e. which degeneracy point is split, varies. Specifically, we

remark that for fixed $N_p$ the number of spectral gaps present decreases on going to larger $N$ and is zero at $N_p/N = 1$ since the number of Casimir values the parabolas can adopt grows with $N$.

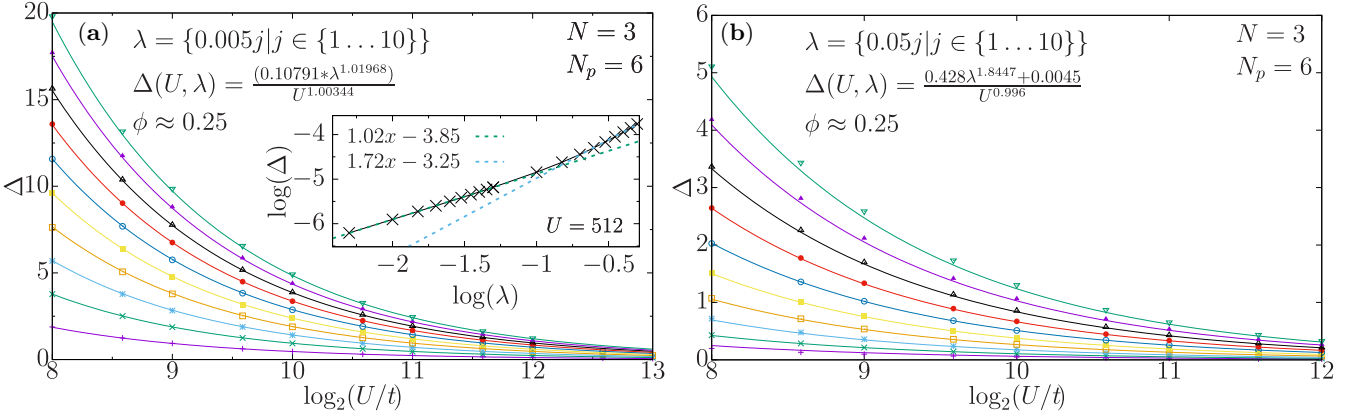

FIG. 13. *Spectral gap behaviour as a function of barrier strength and interaction.* Figures show the gap $\Delta$, at flux $\phi \approx 0.25$ that would correspond to the crossing between parabolas two and three in the ground-state energy landscape, as a function of interaction $U/t$ for $N_p = 6$ particles with SU(3) symmetry at **(a)** weak and **(b)** strong barrier strengths. Inset panel shows the gap versus lambda and interaction fixed to a value $U/t = 512$ in a log-log plot. Two dotted lines show the different behavior as function of $\lambda/t$ and $U/t$. Results obtained with exact diagonalization of the SU(N) Hubbard model with $N_s = 7$ sites. Note that the system is at a sufficiently strong interaction such that $N_p$ peaks are already present in the ground-state.

The scaling of the gap $\Delta$ as a function of the interaction is monotonically decreasing –Fig. 13. To understand the scaling behaviour, we consider the exact Bethe Ansatz equations in the infinitely repulsive limit following the works [57, 60]. In this regime, it was shown that there is a spin-charge decoupling of the Bethe equations such that the system can be described as a composition function of the spinless and SU(N) Heisenberg model, accounting for the charge and spin parts of the system respectively. The energy contribution from the spin-part scales with $1/U$, leading to a full degeneracy of the spin part at $U \to \infty$ where all the gaps close. For weak barrier strengths the scaling is linear as $\Delta(U, \lambda) \approx \lambda/U$ (see Fig. 13(a)), compatible with a perturbative expansion in low-energy descriptions, whilst for stronger ones it becomes non-linear $\Delta(U, \lambda) \approx \lambda^\xi/U$ (see Fig. 13(b)).

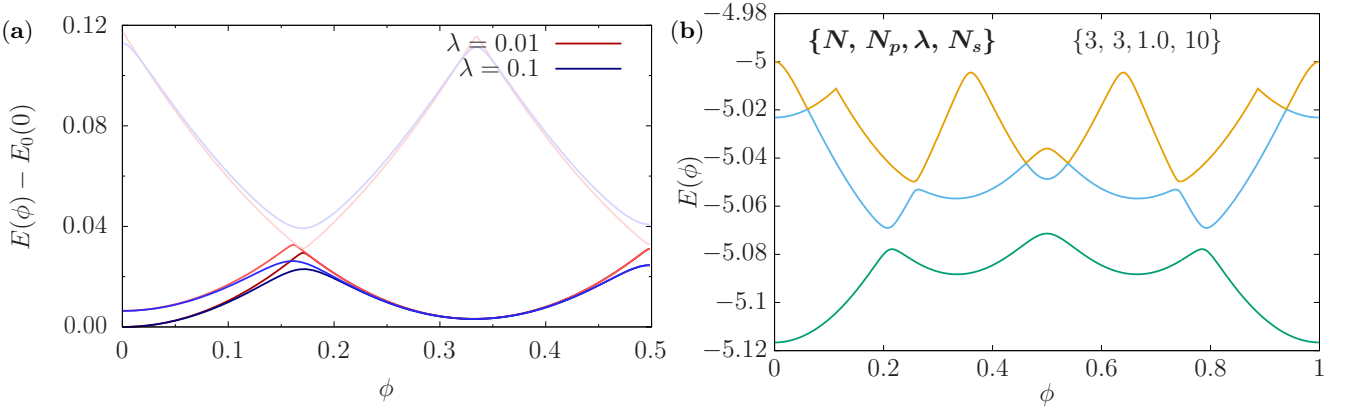

FIG. 14. *Spectral gap opening for SU(N) symmetry breaking.* **(a)** Energy $E(\phi)$ versus flux $\phi$ with a fixed interaction $U/t = 256$ and different barrier strengths (acting only on one species) $\lambda/t$. For any value of the $\lambda/t$ a gap opens up. **(b)** $E(\phi)$ against $\phi$ for fixed $\lambda/t = 1$ where the SU(3) symmetry is broken by choosing different interactions between the components $U_A/t = 10$, $U_B/t = 20$ and $U_C/t = 30$. Results obtained with exact diagonalization of the SU(N) Hubbard model for $N_p = 3$ three-component fermions residing in a ring of $N_s = 10$ sites.

To conclude, we note that by choosing a colour selective barrier, e.g. only applying a barrier for one colour, a gap opens for very small values of the barrier strength $\lambda$. Furthermore, gaps are present everywhere in the ground-state compared to the isotropic barrier case –Fig. 14(a). Similarly, breaking the SU(3) symmetry explicitly by choosing a species-dependent interaction lifts degeneracies opening some of the previously closed gaps –Fig. 14(b). Note that this method of breaking the SU(N) symmetry is only possible for $N > 2$.

 

\end{document}